\documentclass[12pt,preprint]{aastex}

\def\gs{\mathrel{\raise0.35ex\hbox{$\scriptstyle >$}\kern-0.6em
\lower0.40ex\hbox{{$\scriptstyle \sim$}}}}
\def\ls{\mathrel{\raise0.35ex\hbox{$\scriptstyle <$}\kern-0.6em
\lower0.40ex\hbox{{$\scriptstyle \sim$}}}}

\def\spose#1{\hbox to 0pt{#1\hss}}
\def\simlt{\mathrel{\spose{\lower 3pt\hbox{$\mathchar"218$}}
     \raise 2.0pt\hbox{$\mathchar"13C$}}}
\def\simgt{\mathrel{\spose{\lower 3pt\hbox{$\mathchar"218$}}
     \raise 2.0pt\hbox{$\mathchar"13E$}}}

\newcommand{\um}{\,$\mu$m\,}

\newcommand{\spitz}{{\sl Spitzer}}
\newcommand{\zr}[2]{$z$\,$\sim$\,#1\,--\,#2}
\newcommand{\lir}[1]{10$^{#1}$\,$\rm{L}_{\odot}$}
\newcommand{\lsun}{\,$\rm{L}_{\odot}$}

\newcommand{\z}{redshift\ }
\newcommand{\con}{continuum\ }
\newcommand{\ulig}{ULIRGs}

\slugcomment{\it To be submitted to the Astrophysical Journal}

\shorttitle{IRS spectra of high-$z$ ULIRGs }
\shortauthors{Sajina et al.}

\begin{document}

\title{{\sl Spitzer} Mid-Infrared Spectroscopy of  Infrared Luminous Galaxies at $z$\,$\sim$\,2 II: Diagnostics}

\author{Anna Sajina\altaffilmark{1}, Lin Yan\altaffilmark{1}, \\
Lee Armus\altaffilmark{1}, Philip Choi\altaffilmark{2}, Dario Fadda\altaffilmark{3}, George Helou\altaffilmark{1}, Henrik Spoon\altaffilmark{4}}

\altaffiltext{1}{\spitz\ Science Center, California Institute of Technology, Pasadena, CA, 91125}
\altaffiltext{2}{Pomona College, Claremont, CA, 91711}
\altaffiltext{3}{{\sl Herschel} Science Center, California Institute of Technology, Pasadena, CA, 91125}
\altaffiltext{4}{Cornell University, Ithaca, NY, 14853}

\begin{abstract}
We present mid-IR spectral decomposition of a sample of 48 \spitz-selected ULIRGs spanning \zr{1}{3} and likely $L_{\rm{IR}}$\,$\sim$\,$10^{12}$\,--$10^{13}$\lsun.  Our study aims at quantifying the star-formation and AGN processes in these sources which recent results suggest have evolved strongly between the observed epoch and today. To do this, we study the mid-IR contribution of PAH emission, continuum, and extinction. About 3/4 of our sample are continuum- (i.e. AGN) dominated sources, but $\sim$\,60\% of these show PAH emission, suggesting the presence of star-formation activity.  These sources have redder mid-IR colors than typical optically-selected quasars.  About 25\% of our sample have strong PAH emission, but none are likely to be pure starbursts as reflected in their relatively high 5\um\ hot dust continua. However, their steep 30\um\,--\,to\,--\,14\um\ slopes suggest that star-formation might dominate the total infrared luminosity. Six of our $z$\,$\sim$\,2 sources have $EW_{6.2}$\,$\gs$\,0.3\um\ and $L_{14\mu\rm{m}}$\,$\gs$\,$10^{12}$\lsun\ (implying $L_{\rm{IR}}$\,$\gs$\,$10^{13}$\lsun).  At these luminosities, such high $EW_{6.2}$ \ulig\ do not exist in the local Universe. We find a median optical depth at 9.7\um\ of $\langle$\,$\tau_{9.7\mu\rm{m}}$\,$\rangle$\,=\,1.4. This is consistent with local {\sl IRAS}-selected \ulig, but differs from early results on SCUBA-selected $z$\,$\sim$\,2 ULIRGs.  Similar to local ULIRGs about 25\% of our sample show extreme obscuration ($\tau_{9.7\mu\rm{m}}$\,$\gs$\,3) suggesting buried nuclei.  In general, we find that our sources are similar to local ULIRGs, but are an order of magnitude more luminous. It is not clear whether our $z$\,$\sim$\,2 ULIRGs are simply scaled-up versions of local ULIRGs, or subject to fundamentally different physical processes.  
\end{abstract}


\section{Introduction}
The pioneering work from the Infrared Space Observatory, {\sl ISO} \citep[see][for a review]{iso_review} has already shown the power of mid-IR spectroscopy to reveal the dominant power sources
among local infrared luminous galaxies. Diagnostic diagrams separating starburst from AGN dominated sources have been developed from several studies \citep{lutz96,lutz98,genzel98,rigo99,laurent00,tran01,peeters04}. 
Today, this work is being extended using the powerful Infrared Spectrograph (IRS; Houck et al. 2004) 
on the {\sl Spitzer} Space Telescope \citep{werner04}. Its unprecedented sensitivity and efficiency have made it possible to expand the {\sl ISO} work to much larger samples covering wider ranges of physical
properties with better $S/N$ and higher spectral resolution \citep[e.g.][]{jdsmith06,brandl06,schweitzer06,wu06}. High $S/N$ ratio IRS spectra from all these local samples have enabled various teams to explore new parameter space and create new diagnostic tools characterizing different galaxy populations \citep[e.g.][]{dale06, armus_ngc6240,armus06,sturm06,newdiag}.

More importantly, IRS has enabled for the first time the 
mid-IR spectroscopy of IR luminous galaxies at cosmologically
significant distances. Although limited to primarily low resolution data, 
the IRS spectra of high-$z$ samples allow us for the first
time to apply the spectral diagnostic power of the mid-IR
regime to galaxies at $z$\,$\sim$\,1\,--\,3, the epoch when 
LIRGs and \ulig\ \citep[see][for a review]{sm96} make a significant contribution to the 
global averaged luminosity density and to the cosmic infrared
background \citep{lefloch05, dole06,caputi07}. 
Recent studies \citep{houck05,yan05,yan07} have produced such low resolution, mid-IR spectral samples of \spitz-selected $z$\,$\sim$\,2 \ulig. Other IRS samples include near-IR-selected starbursts and X-ray selected AGN \citep{weedman06}, and sub-mm galaxies \citep{karin06,pope_smgs}.  
Physical properties derived from these low-resolution, mid-IR spectra will allow us to put various disparate high-$z$ ULIRG samples in context, and test our understanding of IR luminous populations and their evolution.

This paper is part two of a series of papers which present the results from a \spitz\ spectroscopy survey
of \spitz-selected high-\z\ galaxies \citep{yan05}. The first paper (PAPERI, Yan et al. 2007) 
describes the survey sample, data
reduction, observed mid-IR spectra, and the basic 
properties of the galaxies in the sample, including \z\
distribution, mid-IR luminosities and spectral types. 
This work (PAPERII) presents spectral decomposition, provides diagnostic tools, establishes the physical properties of our $z$\,$\sim$\,2 \spitz\ selected ULIRGs, and discusses related limitations of current IRS spectra. The far-IR and radio properties of our sample are presented in  Sajina et al. (2007, in prep).  

Throughout this paper we adopt the cosmological model with 
$\Omega_{M}$\,=\,0.3, $\Omega_{\Lambda}$\,=\,0.7, and $H_0$\,=\,70.

\section{Data}
\subsection{IRS spectral sample}

PAPERI gives the detailed description of the sample selection, presents the observed spectra and \z measurements.  The full sample has 52 sources  selected
to have 24\um\ flux density brighter than 0.9mJy, and 
very red 24-to-8\um\ and 24-to-R colors\footnote{(1) $R(24,8) \equiv \log_{10}(\nu 
f_{\nu}(24\mu m)/\nu f_{\nu}(8\mu m) \simgt 0.5$; 
(2) $R(24,0.7) \equiv \log_{10}(\nu f_{\nu}(24\mu m)/\nu f_{\nu}(0.7\mu m) 
\simgt 1.0$.}. The IRS low resolution data revealed that 47 of the targets
have measurable \z from the features detected in their
mid-IR spectra, with the majority (33/47\,=\,72\%) at 1.5\,$<$\,$z$\,$<$2\,.7, 
and a smaller fraction (12/47\,=\,28\%) at 0.65\,$<$\,$z$\,$<$\,1.5. An additional source (MIPS279) has a Keck redshift ($z$\,=\,1.23) bringing our sample to 48. 

\subsection{Broadband data}
We supplement the IRS spectra with Infrared Array Camera \cite[IRAC;][]{fazio04}, and Multiband Photometer for \spitz\ \citep[MIPS;][]{rieke04} 70\um\ broadband photometry. 

The IRAC observations were obtained as part of the  \spitz\ Extragalactic First Look Survey (XFLS\footnote{\url{http://ssc.spitzer.caltech.edu/fls/}}) \citep{lacy05}.  The rms noise within a $6^{''}$
diameter aperture is 2.8, 3.2, 15, 14.4$\mu$Jy at 3.6, 4.5, 5.8 and 8.0\um\ respectively. 
 We measured the aperture fluxes of our 48 sources through apertures with a diameter of 7.3" (6 pixels) using the IRAC mosaiced images from Lacy et al. (2005).   We applied aperture corrections of 1.112, 1.113, 1.125, and 1.218 for the 3.6, 4.5, 5.8, and 8.0\um\ fluxes respectively. To obtain the sky variance, we used 12 sky apertures at $\gs$\,10\,pixels away from the source. From our sample, 5 sources are confused by nearby stars/bright galaxies. For these cases, we subtract the contaminant from the images by PSF fitting before performing the aperture photometry on our sources. The sources affected are: MIPS289, MIPS42, MIPS22661, MIPS110 and MIPS227.   

 The MIPS 70\um\ fluxes of our sample come from the combined scan map data \citep{frayer06} and pointed photometry data (PI: P. Choi).  Most of our targets have roughly 700 seconds integration time per pixel.  The 70\um\ fluxes are estimated using a similar method as described for the IRAC bands. The aperture fluxes are measured through 3\,pixel radius apertures. The total fluxes are computed by applying an aperture correction of 2.0.  We also estimate flux errors using 12 sky apertures, and find typical rms flux errors of 1.3\,mJy.  All broad fluxes from $3.6$\,--\,70\um\ are listed in Table~\ref{cont_table}.

\section{Fitting approach}
\subsection{Spectral decomposition model \label{sec_model}}
The mid-IR spectra of a wide range of sources can be thought of as a superposition of starburst-powered PAH component and a predominantly AGN-powered continuum component \citep[see e.g.][]{laurent00,sturm00,tran01}. This PAH-to-contiuum diagnostic has been applied to ULIRG spectra as well \citep{genzel98,tran01}. However, lately the significant obscuration of mid-IR continua of the majority of ULIRGs has been highlighted as an important diagnostic parameter as it can dilute, or even erase, the power source signatures \citep{levenson,newdiag,imanishi}. With this in mind, below we describe our simple phenomenological approach aiming at distinguishing the contributions of PAH emission, continuum emission, and mid-IR obscuration. 

Our spectral fitting consists of two parts. The first is to determine \con and opacity by fitting simultaneously three components:
{\bf 1).} an extinction-free PAH emission spectrum, {\bf 2).}  a power-law continuum, and {\bf 3).} a mid-IR extinction law applied to the continuum.  Because our spectra do not necessarily have the same relative PAH ratio as the assumed PAH template, the second step is to fit {\sl all} available PAH features simultaneously using Lorentz profiles, with the pre-determined continuum fit from the first step. The analytical formula describing the first fitting is as follows. 

\small
\begin{equation}
F_{\nu}=a_{\rm{PAH}}F_{\nu,\rm{PAH}}+a_{c}\nu^{-\alpha}e^{-\tau_{\nu}},
\end{equation}
\normalsize

where $a_{\rm{PAH}}$ and $a_{c}$ are the amplitudes of the two components, $F_{\nu,\rm{PAH}}$ is a PAH template, $\alpha$ is the spectral slope, and $\tau_{\nu}$ is the extinction curve.  The normalized extinction curve is fixed, with $\tau_{9.7}$ being the free parameter. We perform a simultaneous fit on the four free parameters above.

We limit the fitting to the restframe 5\,--\,15\um\ spectral window which is the typical restframe coverage of our IRS spectra. The power-law approximation is assumed to be valid in this limited region, but is expected to break down at shorter and longer wavelengths due to dust sublimation and influence of colder dust components respectively \citep[e.g.][]{armus06}.  

\begin{figure}
\begin{center}
\plottwo{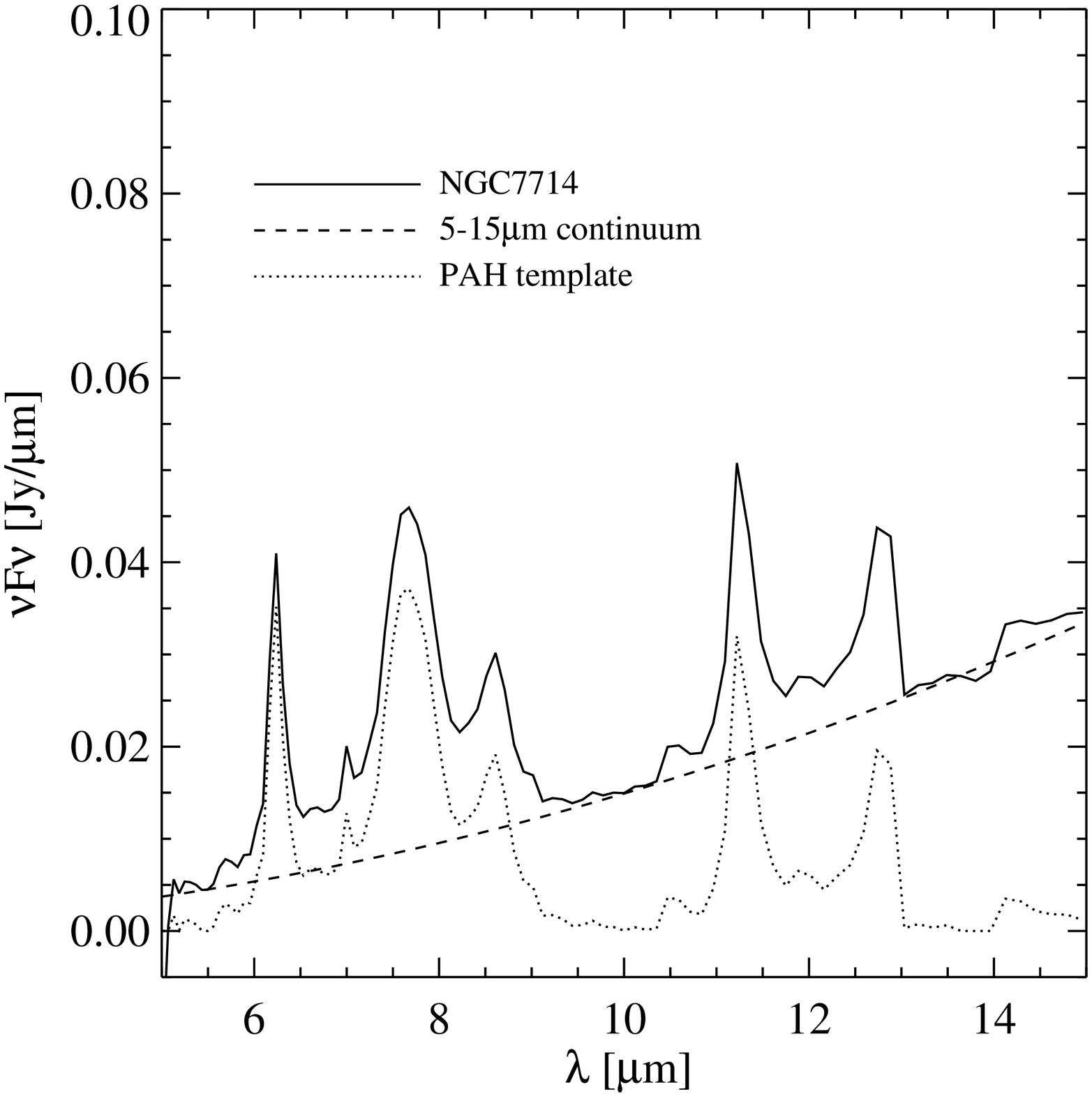}{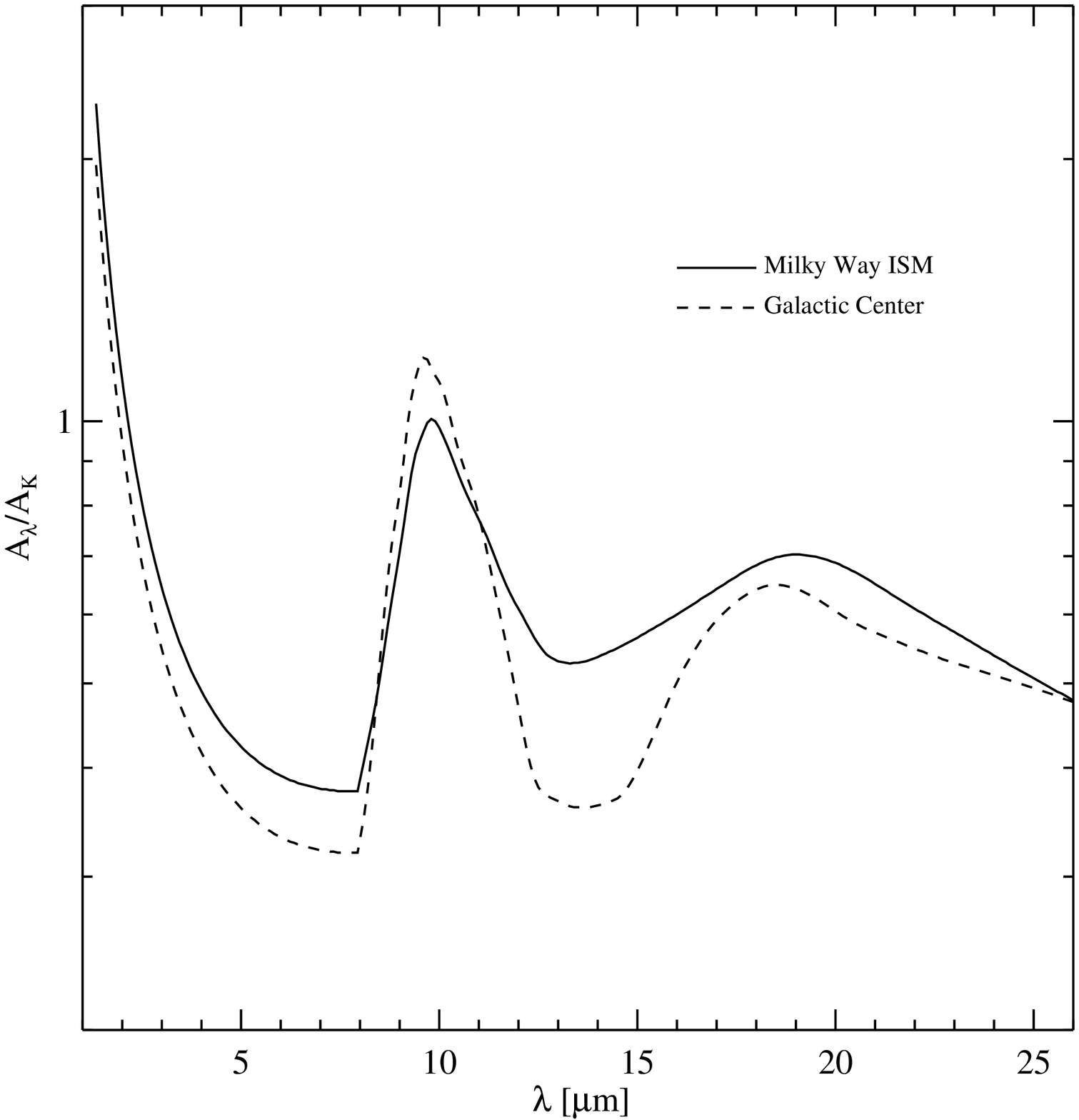}
\end{center}
\caption{{\it Left:} The solid curve is the IRS spectrum of NGC7714. The dashed line is the $\alpha$\,=\,3 power law that best fits the $\sim$\,5\,--\,15\um\ \con without extinction. The PAH template used is the residual (dotted line).  {\it Right:} The Milky Way average (i.e. disk) extinction curve and the Galactic Center extinction curve  \citep{chiar06}.  Here we normalized both to the {\sl K}-band to highlight their differences in the $\sim$\,5\,--\,15\um\ regime of interest.\label{pah_template}}
\end{figure}

Due to the range of redshifts (0.6\,$<$\,$z$\,$<$3.2) of our sources, we see intrinsically different parts of the PAH spectrum.  However, for comparisons within our sample, we need consistent continuum estimates regardless of the rest-frame spectral coverage. This requires adopting a single PAH template, and ignoring PAH variability at this point (see above and \S\,\ref{sec_ew}). The pure PAH spectrum\footnote{This is referred to as a PAH spectrum simply because of the lack of continuum. It is more accurately thought of as "continuum-subtracted starburst spectrum". For simplicity however, we will refer to it as a "PAH template".} ($F_{\rm{PAH}}$) is derived from the IRS, low resolution spectrum of the starburst NGC7714 \citep{brandl06}, where we have subtracted its continuum as shown in Figure~\ref{pah_template} {\it left}.

We tested the effect of our choice of PAH template on the fit results (see \S\,\ref{cont-fit}). We looked at variations in the $\chi^2$ when substituting the NGC7714-based template with the M82 spectrum which intrinsically has much less $\sim$\,5\,--\,15\um\ continuum emission than the average starburst \citep{brandl06}, and therefore we took it  `as is' without subtracting a continuum. We found the variations in the $\chi^2$ to be negligible, i.e. well within the continuum uncertainties. Therefore, for our data, the PAH-continuum decomposition is robust against different PAH template choices. 

\clearpage 
\clearpage

 We choose the Galactic Center (GC) extinction curve \citep{chiar06} as likely closer to the extreme environments in ULIRGs than a standard ISM extinction curve. Figure~\ref{pah_template}{\it right} shows the GC extinction curve compared with the Milky Way disk (MW) extinction curve \citep{chiar06}. 
 
We assume screen extinction\footnote{Since this is the most efficient way to absorb light (compared with mixed and/or clumpy distributions), the optical depths derived in this manner are lower limits.}, which means we ignore radiative transfer effects: absorption and emission along the line of sight in an optically thick medium. 
 
 This approach is similar to some previous studies of local ULIRGs \citep{tran01,lutz03}.
 
 \subsection{PAH profile fitting \label{sec_ew}}

\begin{figure}
\begin{center}
\plotone{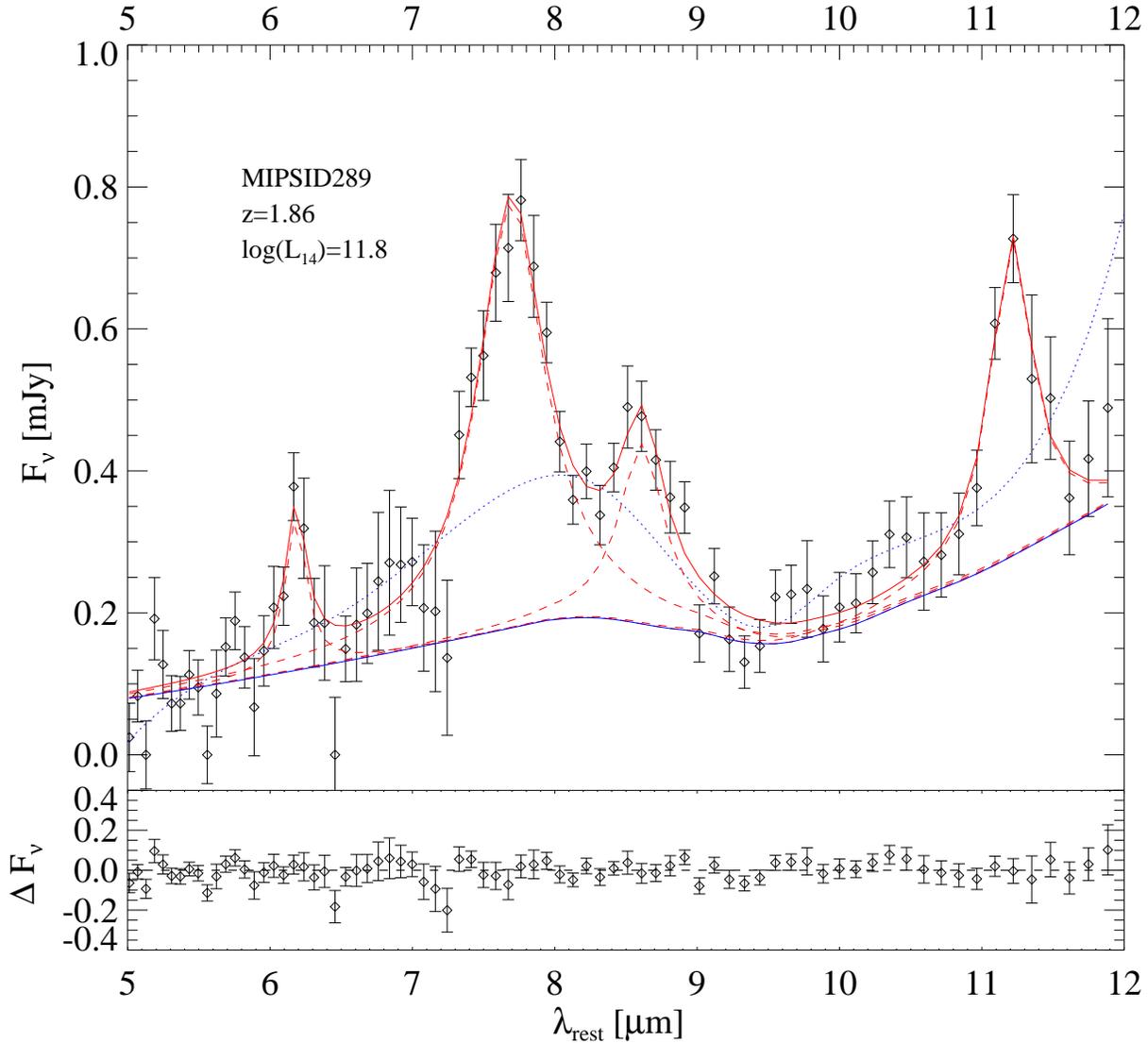}
\end{center}
\caption{An example of our PAH fitting procedure. The best-fit \con is shown in blue, the individual Lorentzian fits are shown as red dashed lines, while the solid red line is the total. For comparison, the dotted curve shows a cubic spline continuum, which misses most of the 7.7\um\ feature as determined using our continuum model \citep[see also,][]{jdsmith06}. The bottom panel shows the residual. \label{pahfit}}
\end{figure}

To accurately measure the strength of observed PAH features, we simultaneously fit a set of Lorentz profiles of the continuum subtracted spectrum at approximately 6.2, 7.7, 8.6, and 11.3\um\footnote{We ignore the 12.6\um\ feature since it is blended with the Ne{\sc ii} 12.8\um\ line.}.  An example is presented in Figure~\ref{pahfit}. Below we give the expressions for line flux and equivalent width, as well as their associated uncertainties. The amplitudes, widths, and exact central wavelengths of the features are all free parameters.  The widths are not fixed because of intrinsic variability such as due to secondary features \citep[e.g.][]{jdsmith06}. The central wavelengths are allowed to vary within a limited range largely due to the redshift uncertainties of our sample (see PAPERI).   
Denoting the flux density as $F_{\lambda}$ and the \con flux density as $F_{c}$, the total flux and its associated uncertainty is determined by Equation~\ref{fformula}. Note that these are the flux densities of each fitted component rather than of the original spectrum.    

\clearpage
\clearpage
\small
\begin{equation}
\label{fformula}
\begin{array}{r}
F=\sum_{\lambda_1}^{\lambda_2}{(F_{\lambda}-F_{c})d\lambda}, \\\\
 \sigma^2(F) = \sum_{\lambda_1}^{\lambda_2}{\left[\sigma(F_{\lambda})d\lambda\right]^2}+\sum_{\lambda_1}^{\lambda_2}{\left[\sigma(F_{c})d\lambda\right]^2},
\end{array}
\end{equation}
\normalsize

where the limits, $\lambda_1$ and $\lambda_2$ are such that they encompass 95\% of the line profile\footnote{For a Lorentz profile this is 5\,$\gamma$, where 1\,$\gamma$ is the half-width encompassing half the power in the line. For comparison, in a  Gauss profile, this corresponds to 2\,$\sigma$. The Lorentz width parameter, $\gamma$, translates into the more familiar Gaussian FWHM via $\gamma$\,$\sim$\,FWHM/1.8}.  

Assuming a constant rms and ignoring the \con uncertainty, the above formula simplifies to the more familiar: 1\,$\sigma$\,=\,$\sqrt{N_{\rm{pix}}}rms\Delta\lambda$. 
Equivalent widths (EW) and their uncertainties are determined by Equation~\ref{ewformula}.

\small
\begin{equation}
\label{ewformula}
\begin{array}{r}
EW = \sum_{\lambda_1}^{\lambda_2}{\frac{F_{\lambda}-F_{c}}{F_{c}}d\lambda}, \\\\
 \sigma^2(EW) = \sum_{\lambda_1}^{\lambda_2}{\left[\frac{\sigma(F_{\lambda})}{F_{c}}d\lambda\right]^2} +\sum_{\lambda_1}^{\lambda_2}{\left[\frac{F_{\lambda}\sigma(F_{c})}{F_{c}^2}d\lambda\right]^2}
\end{array}
\end{equation}   
\normalsize

In cases where the spectral coverage includes a specific line but it is too weak for profile fitting, we estimate 1\,$\sigma$ upper limits by using the uncertainty equations for the flux and equivalent width respectively. This requires adopting a minimum width, $\gamma_{min}$. For 6.2, 7.7 and 11.3 lines respectively this is: 0.1\um, 0.2\um\, and 0.1\um. These values are consistent with the minimum widths found for the detected sources.  

Both the flux density and equivalent width uncertainties are affected by not just the rms of the spectra, but the uncertainty of the continuum level as well (see Equations~\ref{fformula}, \ref{ewformula}). As we discuss in \S~4.1, we use Markov Chain Monte Carlo (MCMC) fitting, which provides us with the probability distributions of the fit parameters. These include both the uncertainty in the data and the uncertainty purely due to the continuum level. We can derive the latter by subtracting the variance of the data from the the MCMC-derived variance by using $\sigma_{tot}^2$\,=\,$\sigma^2(F_{c})$\,+\,$\sigma^2(F_{\lambda})$. This is necessary in order to avoid counting the data rms twice\footnote{Note that this therefore is different from the uncertainty in the continuum luminosities, which we estimate from the marginalized probability distribution as is, since in that case we want to include the data rms as well (see \S\,\ref{cont-fit}).}. This is a non-negligible effect as we find that the continuum error dominates at 7.7\um\ and 11.3\um, and for $\sim$\,1/2 of the sources at 6.2\um. 

\clearpage
\begin{figure}
\plotone{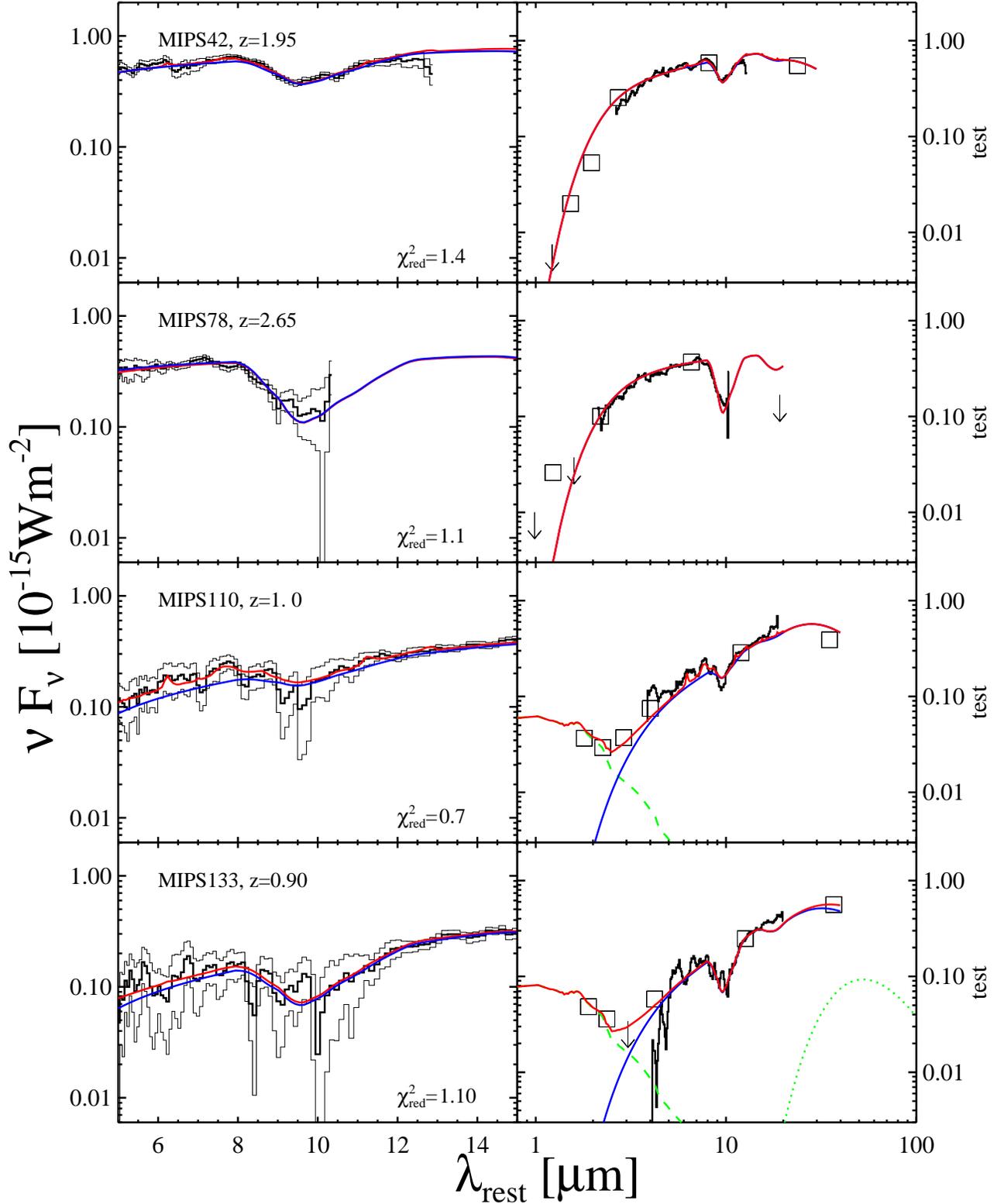}
\caption{
{\it Left:} The IRS spectra ($\pm$\,1$\sigma$) with the best-fit continuum (blue) and continuum+PAH (red). {\it Right:} Including the IRAC and MIPS data (squares).  Overlaid is  our model plus the host galaxy (dashed green) and cold dust excess.   
\label{mirspec}}
\end{figure}
\clearpage
{\plotone{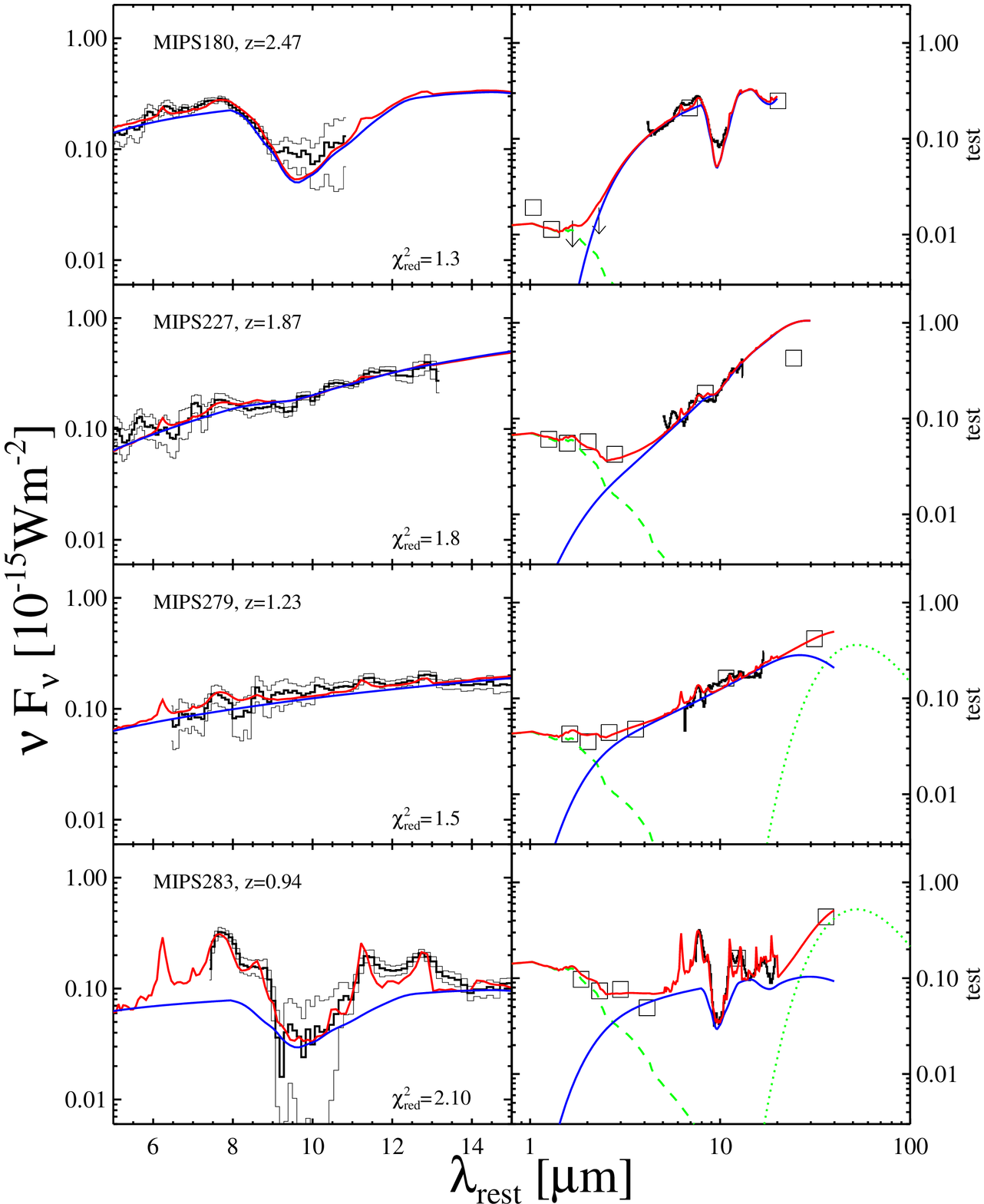}}
\centerline{Fig. 3. --- Continued.}
\clearpage
{\plotone{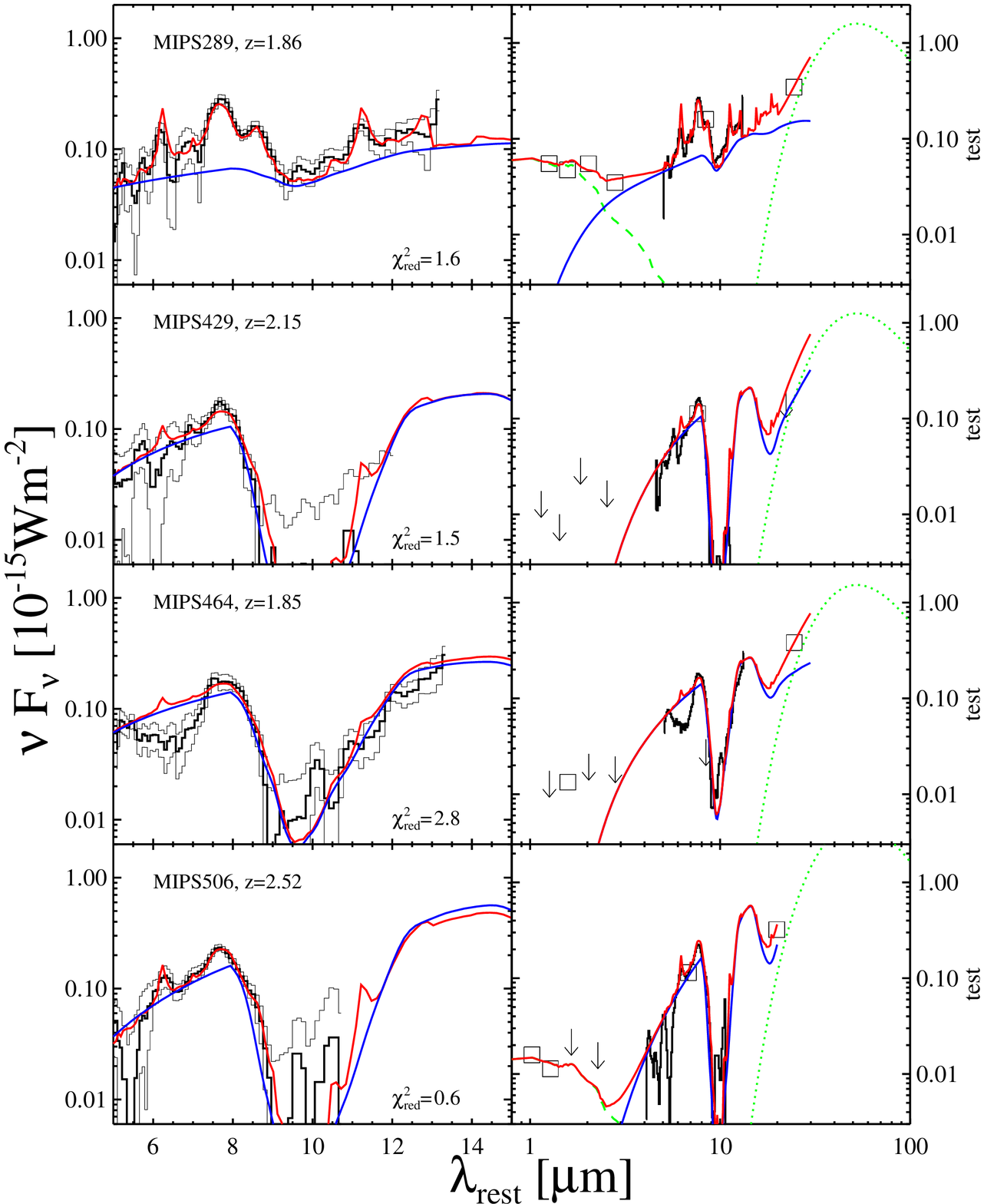}}
\centerline{Fig. 3. --- Continued.}
\clearpage
{\plotone{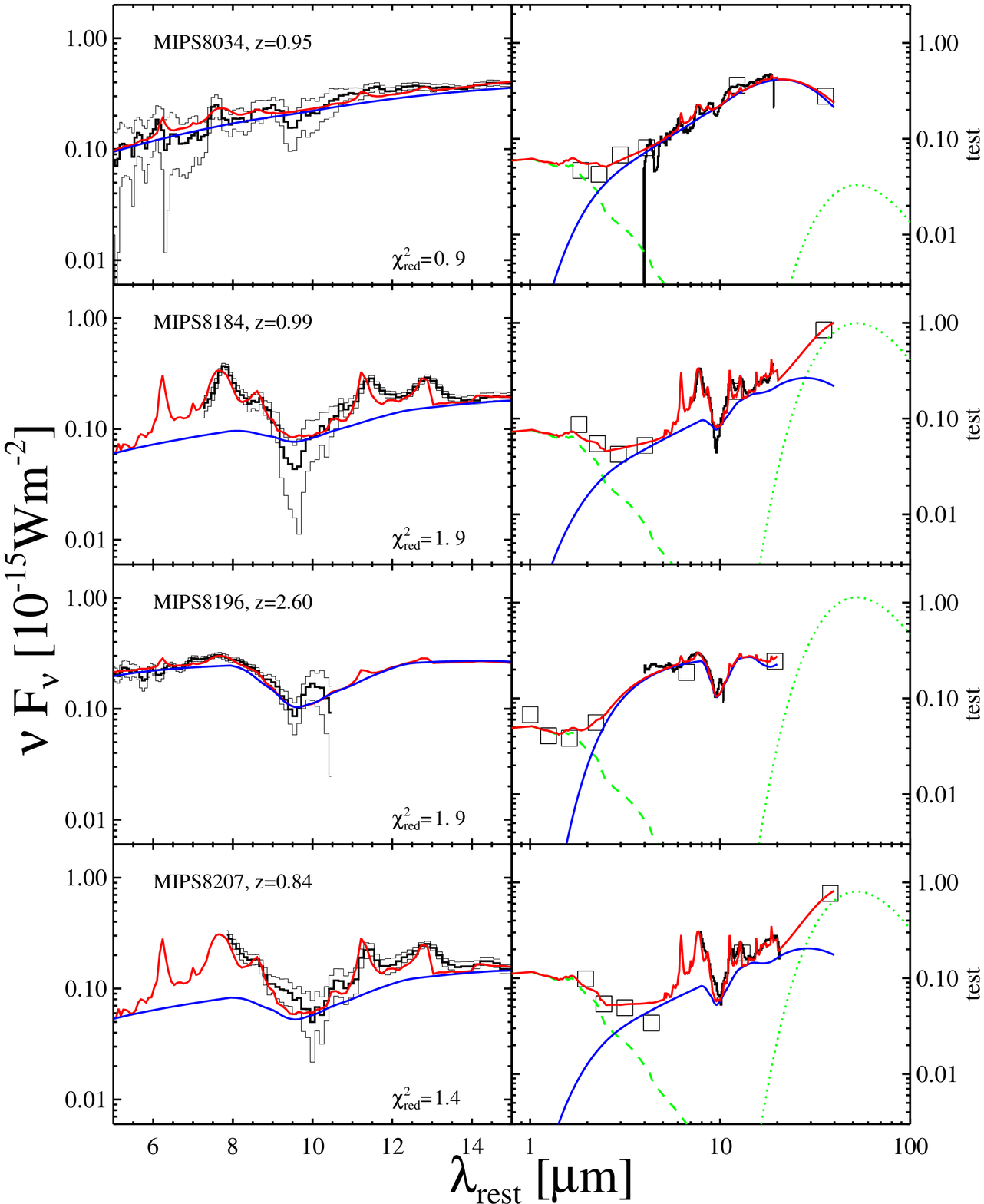}}
\centerline{Fig. 3. --- Continued.}
\clearpage
{\plotone{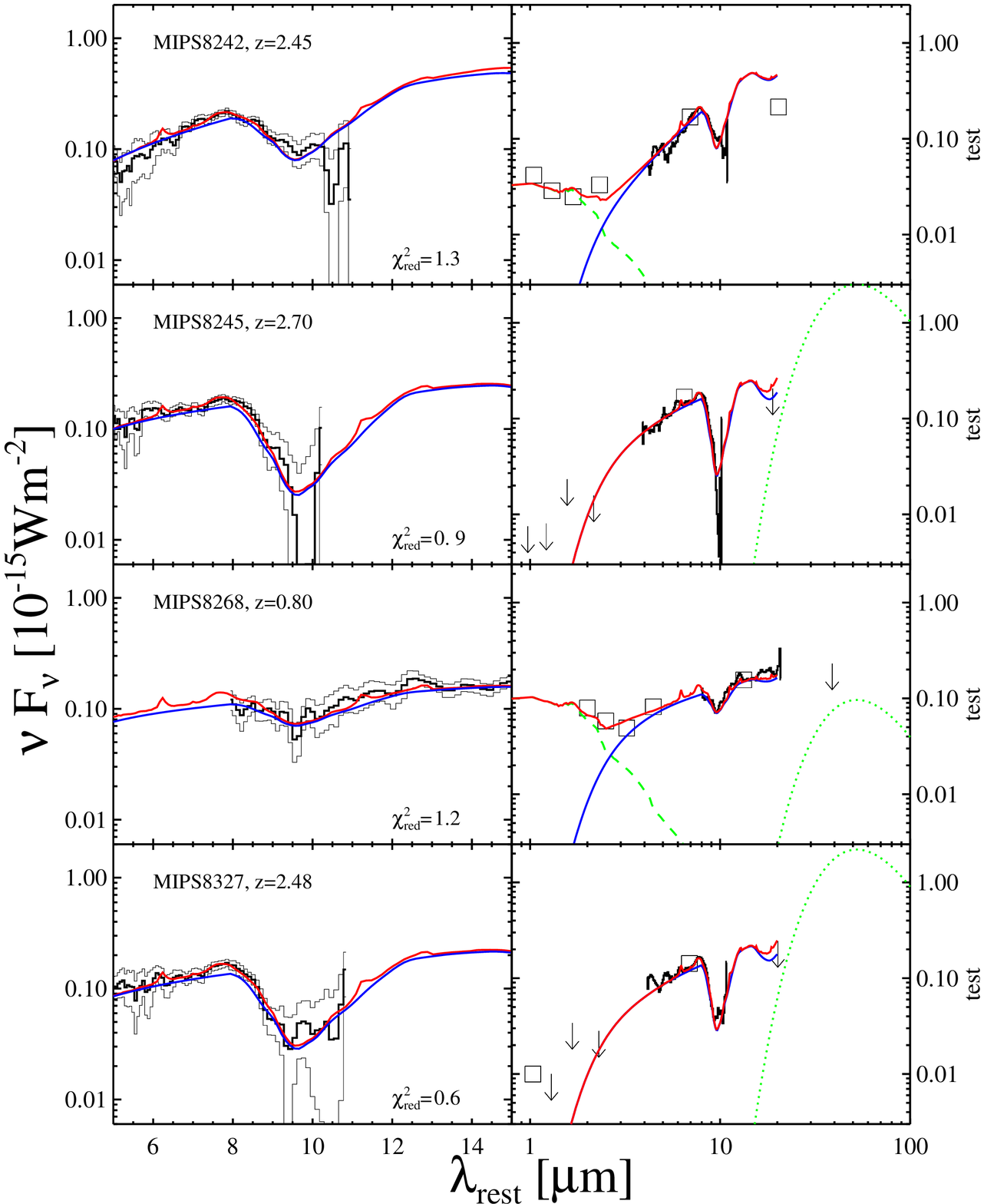}}
\centerline{Fig. 3. --- Continued.}
\clearpage
{\plotone{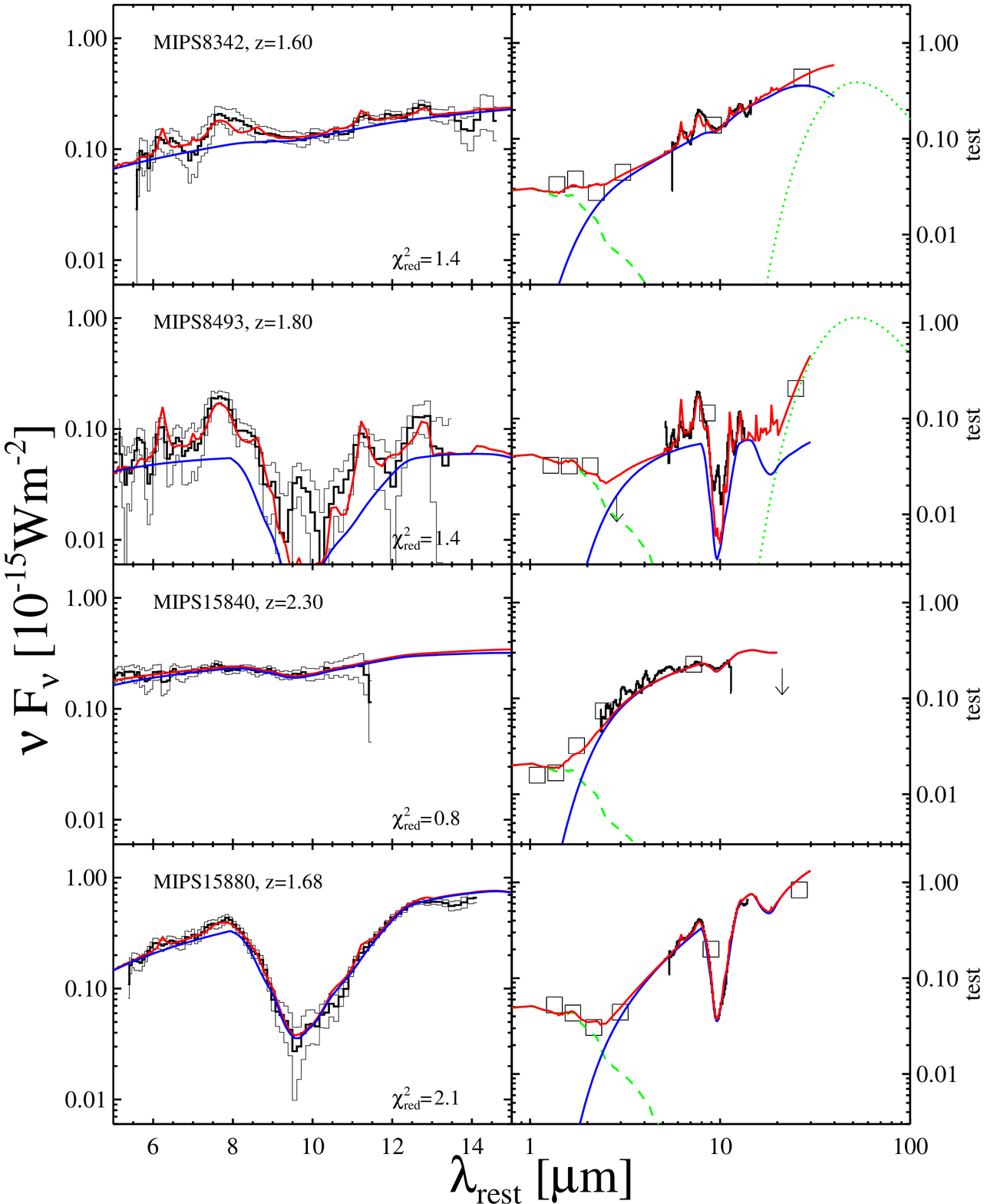}}
\centerline{Fig. 3. --- Continued.}
\clearpage
{\plotone{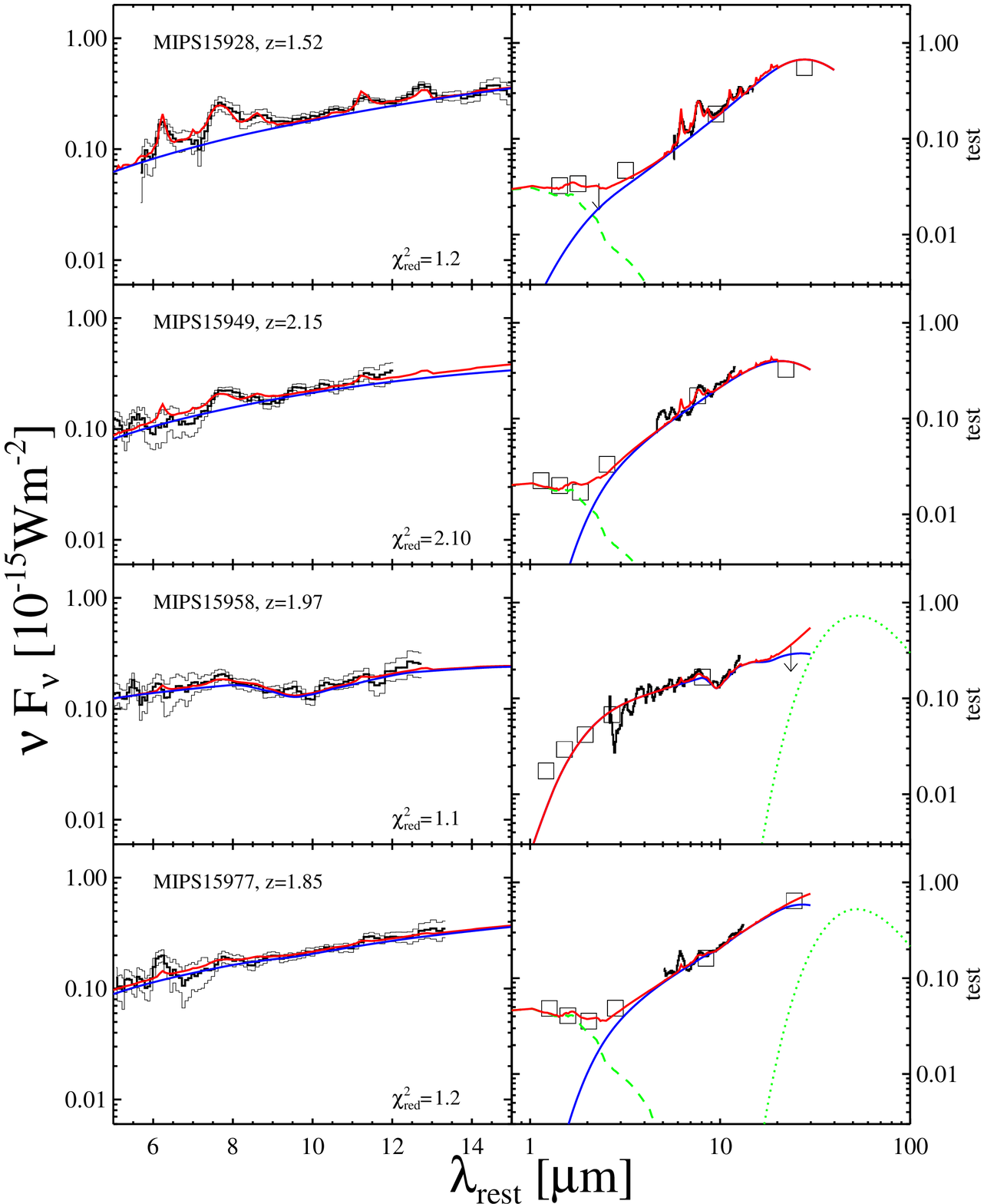}}
\centerline{Fig. 3. --- Continued.}
\clearpage
{\plotone{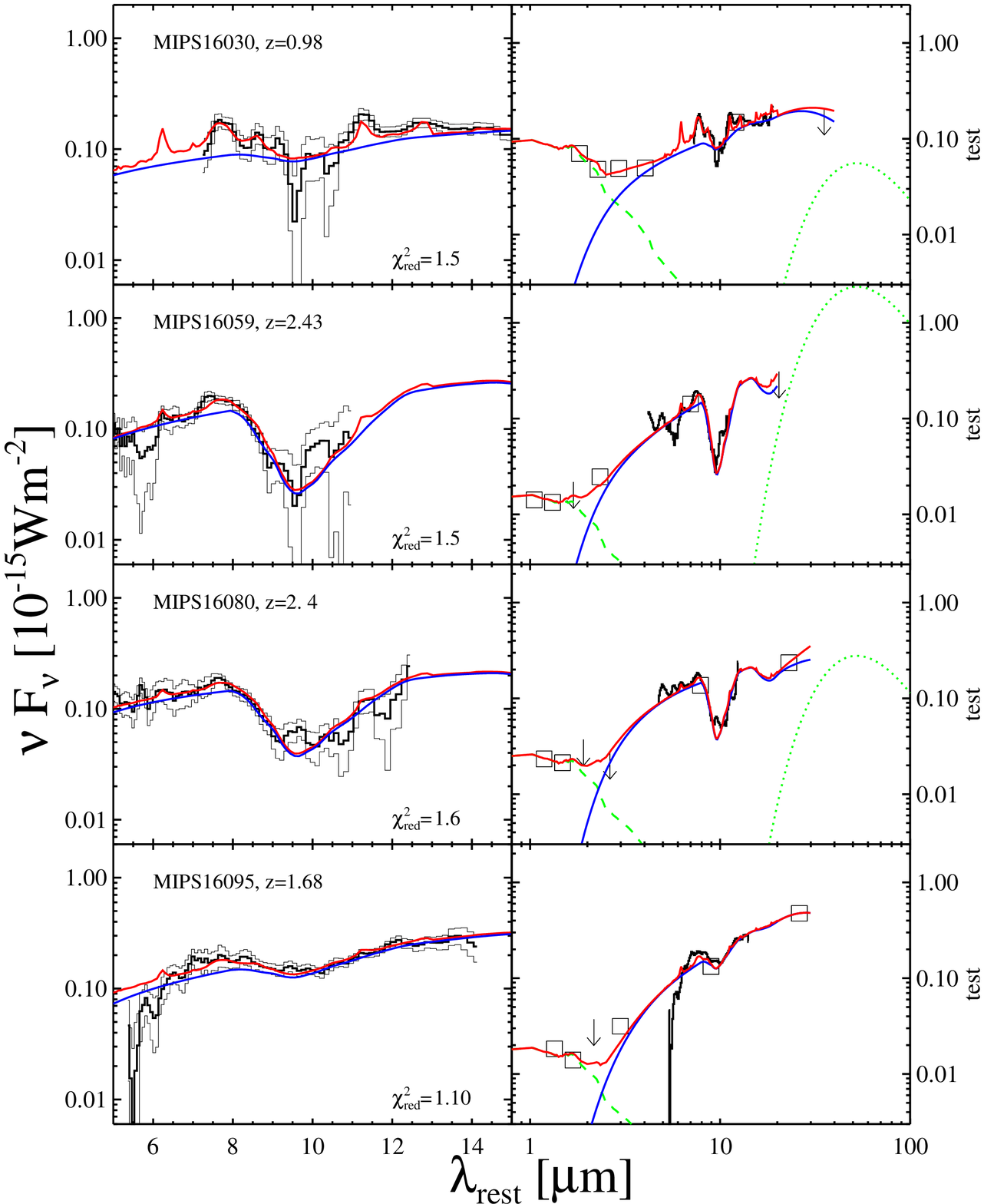}}
\centerline{Fig. 3. --- Continued.}
\clearpage
{\plotone{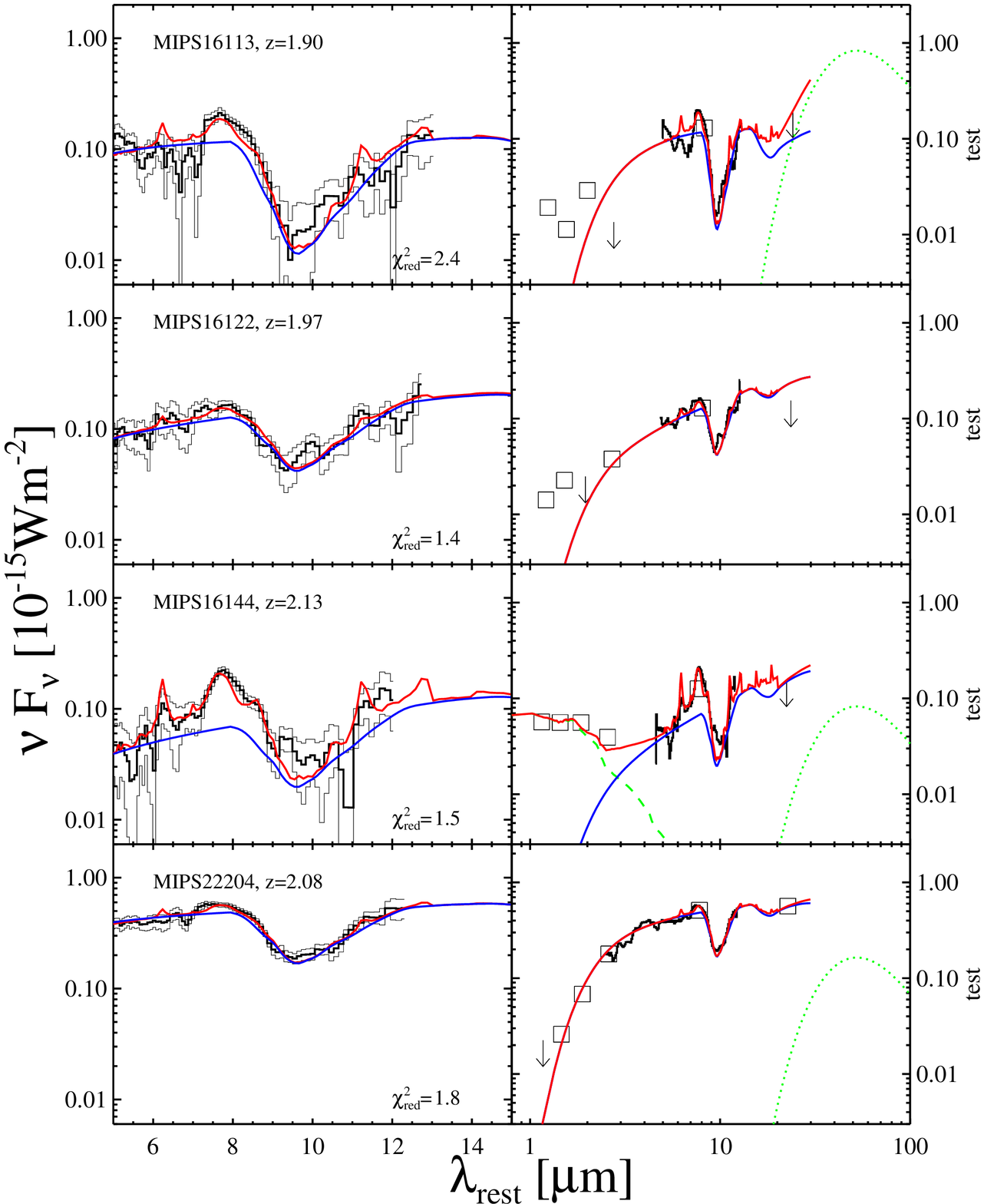}}
\centerline{Fig. 3. --- Continued.}
\clearpage
{\plotone{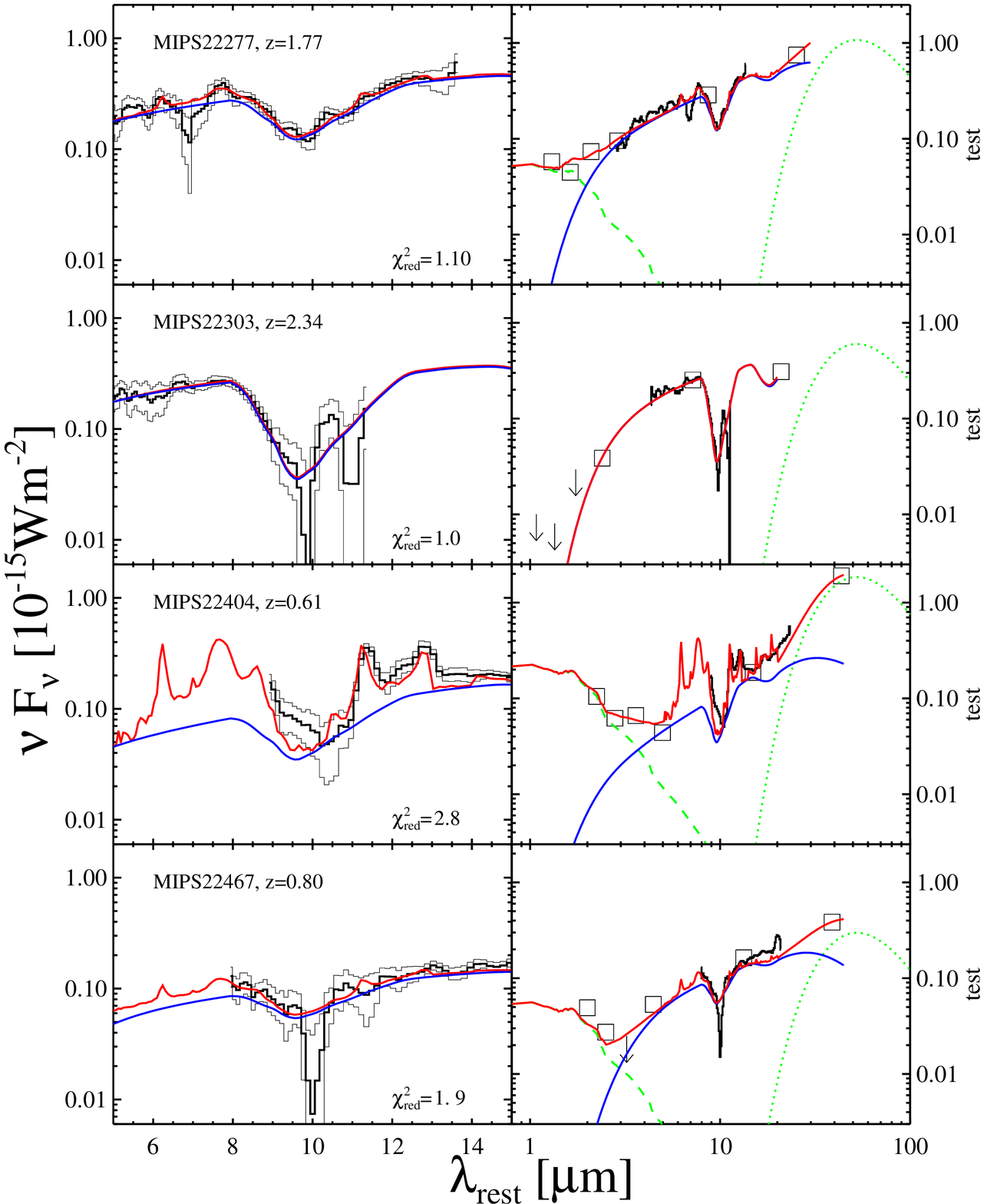}}
\centerline{Fig. 3. --- Continued.}
\clearpage
{\plotone{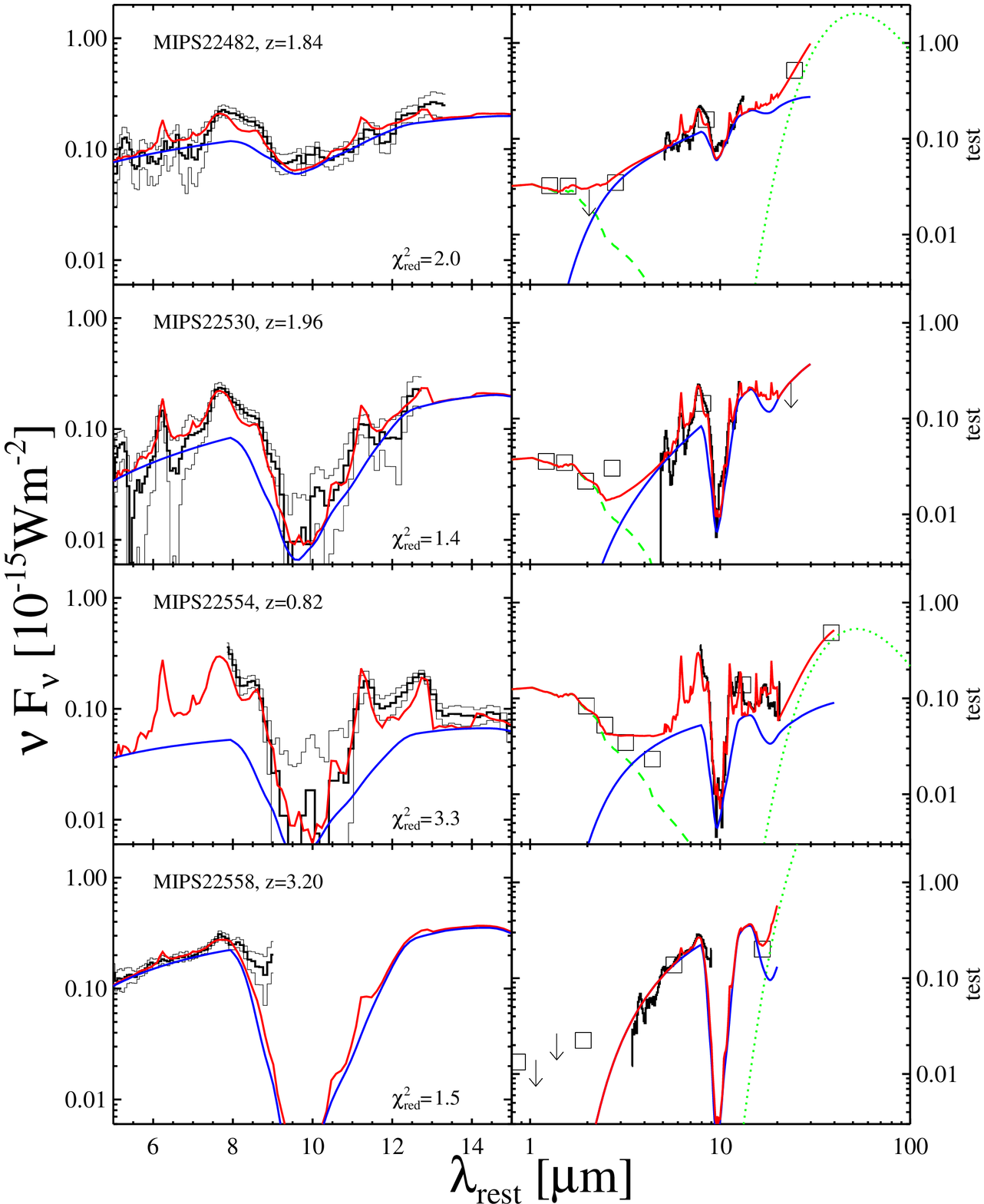}}
\centerline{Fig. 3. --- Continued.}
\clearpage
{\plotone{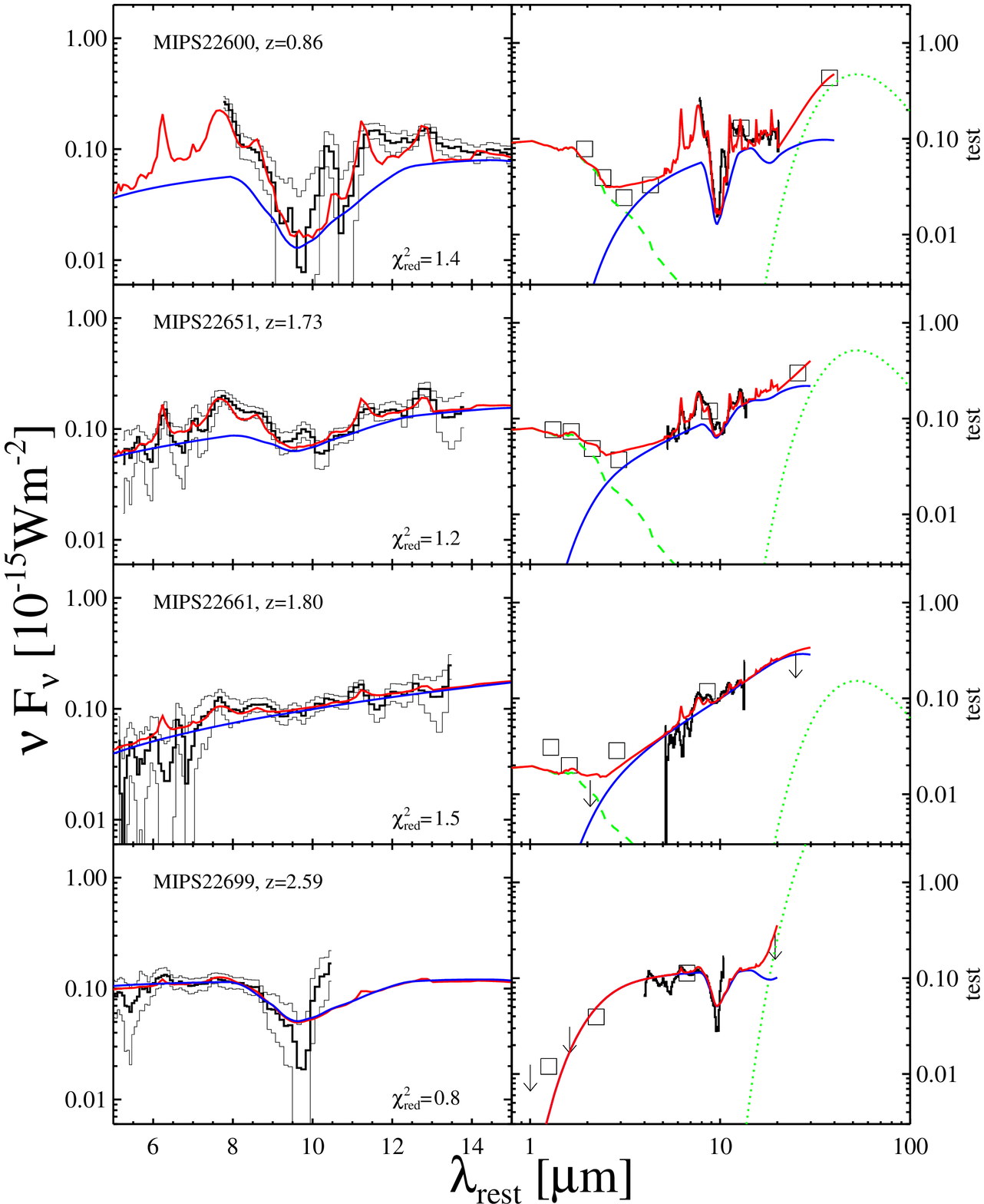}}
\centerline{Fig. 3. --- Continued.}
\clearpage

\section{Fitting results}
  
\subsection{Spectral Decomposition \label{cont-fit}}

We fit the model described in \S\,\ref{sec_model} using Markov Chain Monte Carlo (MCMC; see Sajina et al. 2006 for a description and references).  The left-hand panels of Fig.~\ref{mirspec} show the 5\,--\,15\um\ spectra (over which the fit was performed)  overlaid with the best-fit model.  For the $z$\,$>$\,2.5 sources, we include the MIPS70\um\ point in the fit to help constrain the $\lambda$\,$>$\,13\um\ continuum. A number of typically weaker features\footnote{The start and end couple of microns of the spectra frequently show apparently strong, but likely spurious features.} are likely to be present in our data, but are unaccounted for by the model. These include the neon lines, especially at 12.8\um; absorption features at 6.0\um, 6.9\um\ and 7.3\um\ due to water ice and hydrocarbons \citep[see][]{spoon02,spoon03}, as well as the 9.7\um\ H$_2$ emission feature \citep{higdon06}.  With this caveat, we find that our simple model provides reasonable fits to the spectra (the fit for MIPS464 was forced to go through the 5.5\um\ continuum due to the strong 6\um\ absorption -- see appendix).  The best-fit continuum parameters are given in Table~1. The uncertainties are determined from the probability distributions sampled by the MCMC. 

Our model fitting is performed on the $\sim$\,5\,--\,15\um\ spectra alone as stated above. However, the availability of the IRAC and MIPS70\um\ data allow us access to additional diagnostics as described in \S\,\ref{sec_excess}. Therefore in the right-hand panel of Figure~\ref{mirspec}, we show the IRS spectra along with the available IRAC3.6\um, 4.5\um, 5.8\um, and 8.0\um\ as well as MIPS70\um\ broadband points.  We overlay the best-fit model shown in the left-hand panel, but with one important modification. Since the power-law approximation cannot physically extend to much shorter and longer wavelengths, we truncate it below 5\um\ and above 15\um\footnote{This is done by assuming the power-law is a collection of blackbodies and setting T$_{\rm{min}}$\,=\,600\,K and T$_{\rm{max}}$\,=\,1000\,--\,1500\,K.}.  It is expected that the IRAC points would  be sensitive to host galaxy stellar light (if present), while the 70\um\ points ($\lambda_{\rm{rest}}$\,$\gs$\,20\um) would be sensitive to cooler dust emission (if present).  The data do not allow sophisticated study of either of these components, and hence we characterize each with a single template (see \S\,\ref{sec_excess} for details).  For each of these two components a simple linear fit is performed to find the best-fit amplitude for the corresponding template. In the right-hand panels of Figure~\ref{mirspec}, we show the best-fit host galaxy component (dashed green curve), and the best-fit thermal dust component (dotted green curve).  The red curve shows the best-fit mid-IR model with these two components added to it (see \S\,\ref{sec_excess} for a discussion). 

\subsection{Mid-IR \con slope and opacity \label{sec_cont}}

\begin{figure}
\plotone{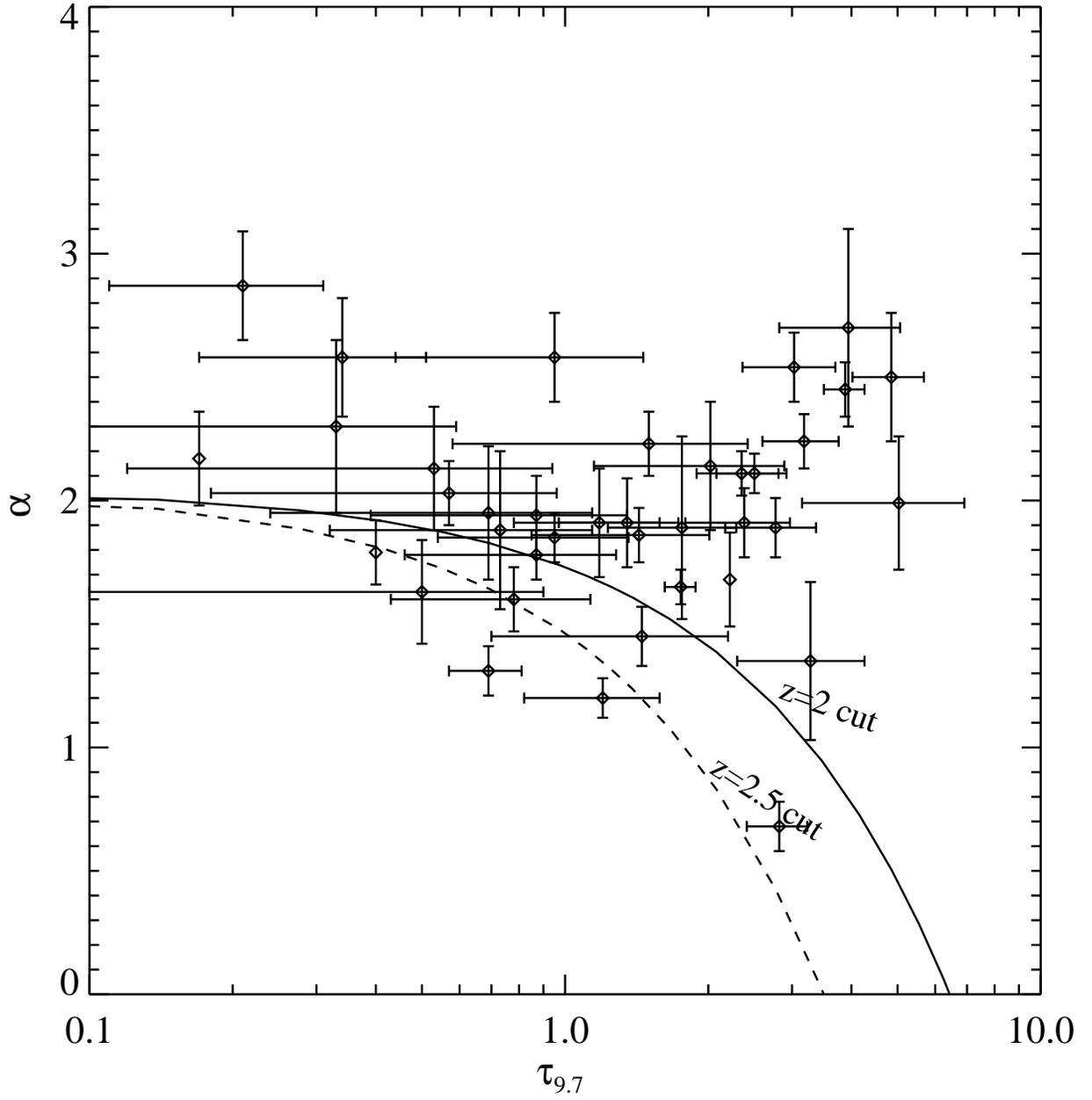}
\caption{The \con parameters $\alpha$  and $\tau$.  The curves show the effect of our $S_{24}/S_{8}$ color selection for $z$\,=\,2 and $z$\,=\,2.5 as labeled.  
\label{cont_diff} }
\end{figure}

Figure~\ref{cont_diff} shows the distribution of optical depth vs. \con slope for our sample. There is a notable lack of sources with low $\tau$ and flat spectra (i.e. $\alpha$\,$\sim$\,1). This is due to one of our selection cuts, $\log(S_{24}/S_{8})$\,$>$\,0.5 (see PAPERI). The effect of this cut as a function of optical depth is shown in Figure~\ref{cont_diff} for $z$\,=\,2 and $z$\,=\,2.5. The median parameter values are: $\langle$\,$\alpha$\,$\rangle$\,=2.0 and $\langle$\,$\tau_{9.7}$\,$\rangle$\,=1.4.  
\clearpage
\clearpage
Our optical depth measurements assume an extinction curve acting on a power-law continuum. Another common measure is the apparent depth of the silicate feature compared with the continuum defined by the 5\um\ and 15\um\ points \citep[e.g.][]{brandl06,newdiag}.  Given a fixed extinction curve, the two quantities have a simple scaling relation. In the case of the Galactic Center extinction curve we adopt, this relation is simply: $\tau_{9.7}$\,=\,1.4\,$\times$\,$\tau_{Si, \rm{obs}}$. 

\subsection{Monochromatic luminosities \label{sec_excess}}
We use the best-fit continua of our IRS spectra to derive monochromatic luminosities ($\nu L_{\nu}$) in two pure continuum regions.  We pick 5.8\um\ due to its sampling the hot dust continuum just shortward of the 6.2\um\ PAH feature, and 14\um\ for its sampling the continuum between the 9.7\um\ and 18\um\ SiO absorption features (it is a minimum in the GC extinction curve see Figure~\ref{pah_template}). We read these luminosities from the best-fit continuum fits not directly from the data. This is necessary because of variable rest-frame spectral coverage, noise spikes at the beginning and end of spectra. This also means that observed absorption features (e.g. MIPS464) not present in the model, do not affect these luminosities. 

The IRAC and MIPS70\um\ photometric points allow us to sample the SEDs of our sources outside of the regions covered by the IRS spectra (see \S\,\ref{cont-fit}). Here we quantify the emission shortward and longward of the available spectra by deriving the monochromatic 1.6\um\ and 30\um\ emission wherever possible. As discussed in \S\,\ref{cont-fit}, the IRAC points (rest-frame $\sim$\,1\,--\,3\um) are likely to be sensitive to direct stellar emission from the host galaxy, which has a characteristic bump at 1.6\um. We only accept that we are likely seeing host galaxy light  if: 1) the source is detected in both 3.6\um\ and 4.5\um, and 2) if the spectral shape changes color from the dust continuum (e.g. yes for MIPS15949, but no for MIPS15958). These data do not allow us to explore the properties of the stellar populations. All we can address is the amplitude of the host galaxy emission whenever present.  To do so we subtract from the IRAC points the mid-IR continuum (truncated below $\sim$\,5\um\ as described in \S\,\ref{cont-fit}), and  perform a simple linear fit using a 10\,Gyr-old stellar population template \citep{pegase}.  The choice of template here is irrelevant, as it merely allows us to find the amplitude of the host galaxy emission (quantified via the monochromatic 1.6\um\ luminosity).

On the other hand at $z$\,$<$\,1.6, the MIPS70\um\ point samples restframe $\sim$\,30\um.  If it is present, this traces the colder dust emission dominating the typical star-forming galaxy's SED. We therefore fit a 50\,K thermal emission to the MIPS70\um point, whenever it is in excess of the truncated power-law continuum. We emphasize that, as in the case of the stellar population, the current data do not allow us to determine the SED peak (and hence dust temperature).  Therefore,  the 50\,K blackbody is used {\it only} as a spectral templates allowing us to derive the monochromatic 30\um\ luminosity.  

The broadband-derived 1.6\um\ and 30\um\ luminosities, along with the 5.8\um\ and 14\um\ luminosities derived from the IRS spectral fitting itself are shown in Table~\ref{lums_table}.

\subsection{PAH features strength}

To determine the PAH features strength, we start by subtracting the best-fit continuum from the spectra. We then perform the simultaneous Lorentz profile fits to the main features as described in \S\,\ref{sec_ew}.  

The PAH equivalent widths and fluxes are tabulated in Table~\ref{pah_table}, where the values less than 2$\sigma$ (this includes both the data and continuum uncertainties) are shown as upper limits (sources without spectral coverage have no value). We leave off the 8.6\um\ feature as it is too confused with the 7.7\um\ feature and affected by the continuum uncertainties (due to the Si feature). We only include it in the fitting primarily for the sake of improving the 7.7\um\ feature strength estimates. 

Out of a total of 48 sources, 38 have data at 6.2\um\ of which 2 have the 6.2\um\ PAH feature detected at $>$\,3\,$\sigma$ (9 at $>$\,2\,$\sigma$). Out of 46 sources with spectral coverage, 24 have their 7.7\um\ feature detected at $>$\,3\,$\sigma$ (36 at $>$\,2\,$\sigma$).  For the 11.3\um\ feature, out of 37 sources with data 12 are detected at $>$\,3\,$\sigma$ (20 at $>$\,2\,$\sigma$). Although the uncertainties vary within the sample, the minimum detectable equivalent widths are roughly EW$_{\rm{6.2}}$\,$\gs$\,0.2\um, EW$_{\rm{7.7}}$\,$\gs$\,0.3\um, and EW$_{\rm{11.3}}$\,$\gs$\,0.3\um. Care must be taken when comparing our EW measurements with those obtained using different approaches. The difference in 6.2\um\ feature equivalent width measurements are well within $\sim$\,50\% between different fitting approaches, however our 7.7\um\ equivalent widths are about a factor of 4 larger than those obtained by using a cubic spline continuum \citep[e.g.][]{brandl06,armus06}. 

The luminosity of the strongest PAH feature, $L_{7.7}$ is given in Table~\ref{lums_table}, along with the key monochromatic luminosities tracing the SEDs  discussed in \S\,\ref{sec_excess}. 
\clearpage
\begin{figure}
\begin{center}
\plottwo{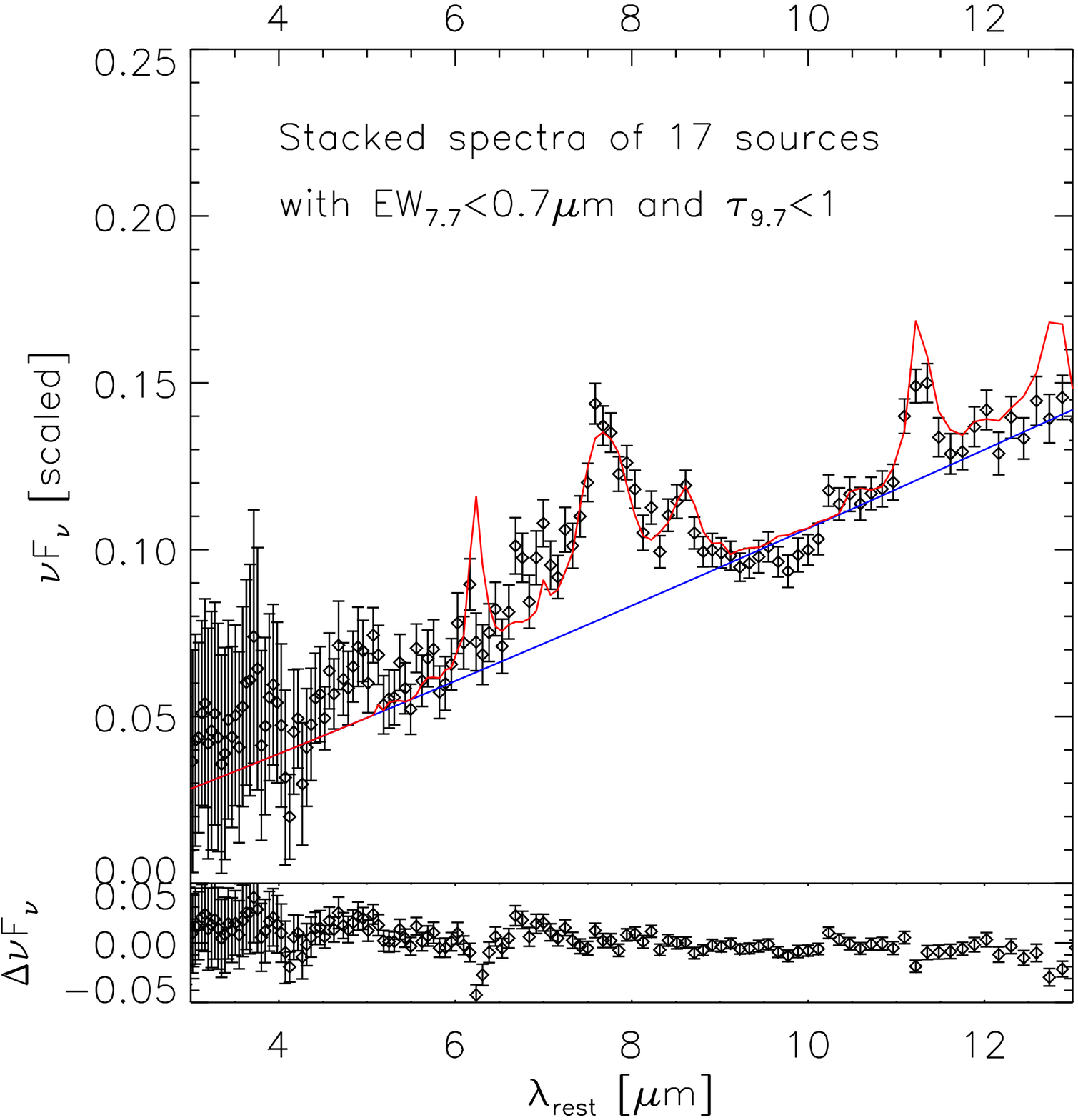}{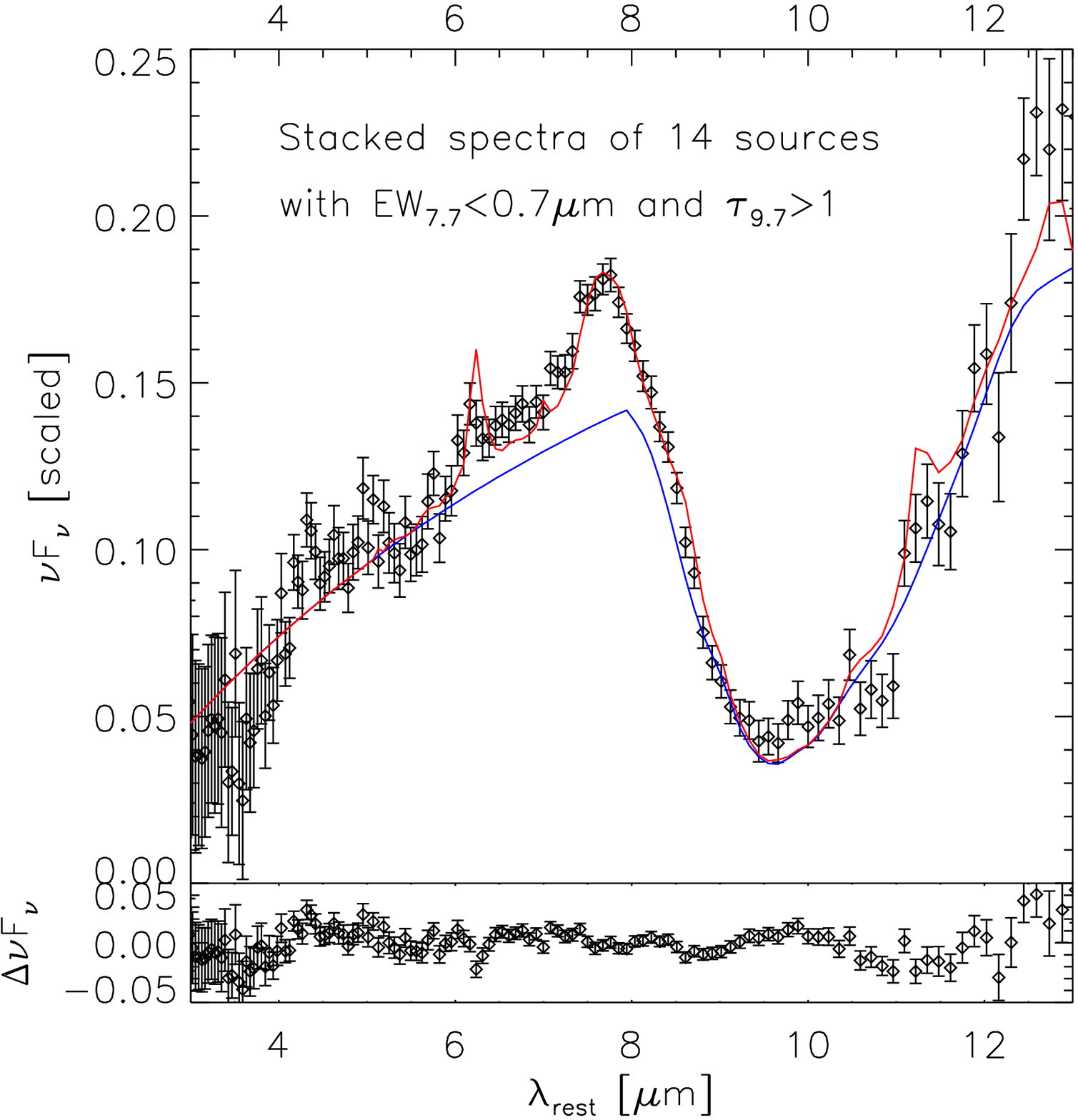}
\end{center}
\caption{The stacked spectra of weak-PAH sources divided into optically-thin ($\tau_{9.7}$\,$<$\,1) and optically-thick ($\tau_{9.7}$\,$>$\,1). Note that a simple model with our PAH template added to a \con (in blue) is a reasonable fit to the data.  The power-law slope is 2.1 for the low-$\tau$ spectrum,  and 1.7 for the high-$\tau$ spectrum. A $\tau_{9.7}$\,=\,2.0 extinction is applied to the later.  \label{weakpahstacks}}
\end{figure}
\clearpage
\subsubsection{Stacking weak-PAH spectra \label{sec_pah}}
In Figure~\ref{weakpahstacks}, we show the stacked spectra of the weakest-PAH sources (i.e. EW7.7\,$<$\,0.7\um), separated into optically-thin ($\tau_{9.7}$\,$<$\,1) and optically-thick ($\tau_{9.7}$\,$>$\,1) sources.  Individual spectra were first scaled to have the same total power (i.e. $\sum{\nu F_{\nu}}$) which is less biased than normalizing at a particular wavelength. The stacked spectra are the weighted means of these scaled spectra. These stacked spectra clearly show that the PAH+continuum interpretation is valid even if this is not always obvious in individual spectra. As discussed in \S\,\ref{cont-fit}, other features are likely to be present in the spectra, which could lead to misidentification of features and therefore wrong redshifts.  Figure~\ref{weakpahstacks} suggests that this is unlikely for the vast majority of the sample. 

The stacked spectra of the high-$\tau$, weak-PAH sources highlights that the distorted hump at $\sim$\,8\um\ is easily explained by a standard PAH emission spectrum superimposed on a deeply obscured continuum. The presence of the 6.2\um\ feature confirms this. However, we have excluded MIPS429 and MIPS464 due to their exceptionally strong $\sim$\,6\um\ absorption features likely due to water ice and/or hydrocarbons (see Appendix~A). The average of the remaining obscured weak-PAH sources still shows  a weaker 6.2\um\ feature compared with our template, which suggest that the same absorption might be present in these sources as well. On the other hand, the weak-PAH low-$\tau$ sources also show a 6.2\um\ PAH feature that is  weaker relative to our PAH template.  This may indicate smearing due to redshift uncertainties, PAH obscuration, or variations in the intrinsic PAH features strength. 

\section{Starburst-AGN diagnostics}
\subsection{The PAH features strength \label{pahdiag}}
\subsubsection{7.7\um\ PAH feature \label{ew7}}
We focus on the 7.7\um\ feature as it is usually the strongest and this spectral region is observed for all of our sample.

\begin{figure}
\begin{center}
\plotone{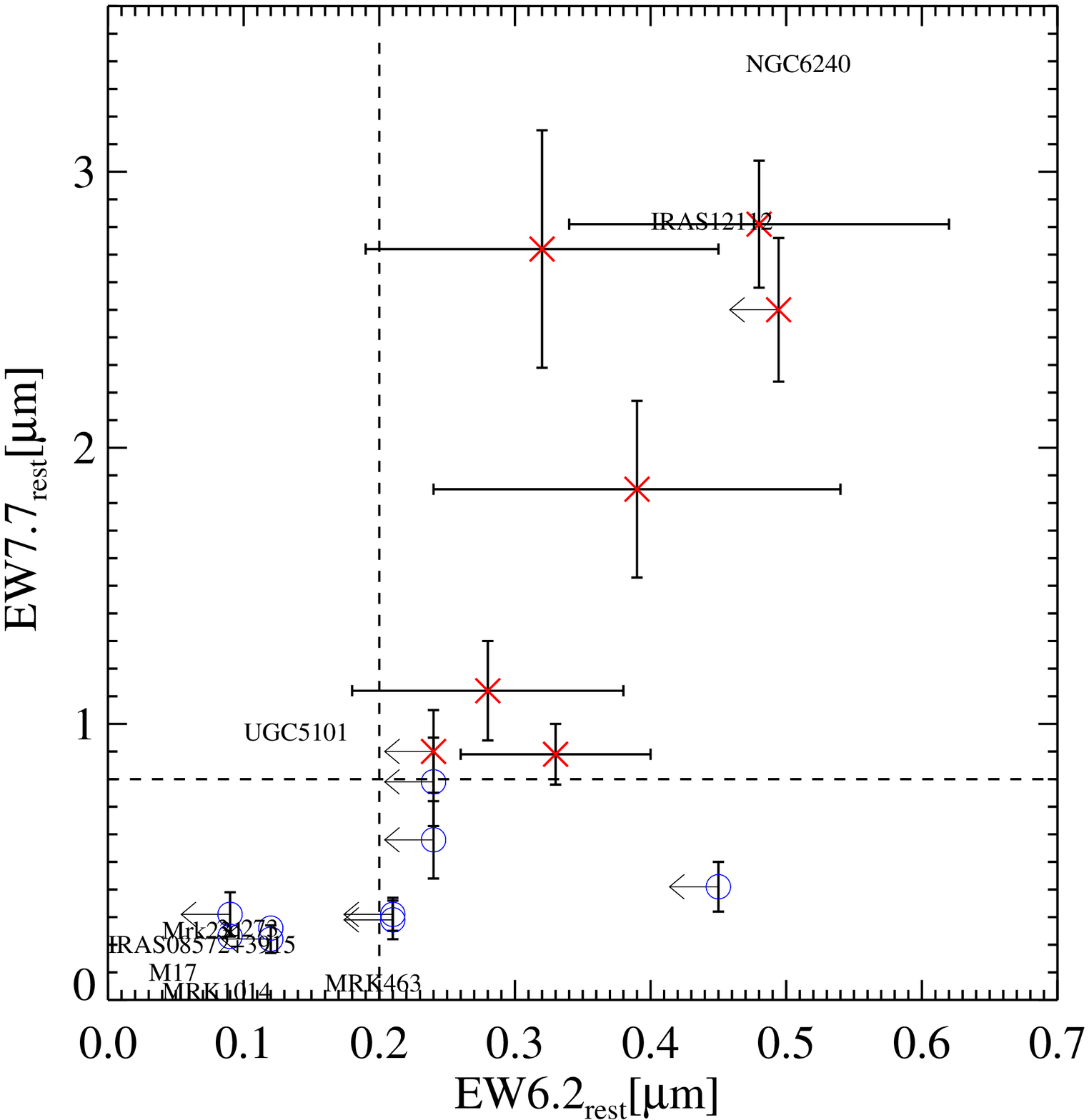}
\end{center}
\caption{Here we show the sources with EW7.7\um\,$>$\,3\,$\sigma$, where data are available at 6.2\um\ as well.  The 6.2\um\ detection limit is EW6.2\um\,$\gs$\,0.2\um,  corresponding to EW7.7\um\,$>$\,0.8\um. Both limits are shown as dashed lines. We will use the 7.7\um\ equivalent widths in the rest of the paper as it is available for most of our sources. From now on, red crosses denote PAH-rich sources (EW7.7\um\,$\geq$\,0.8\um), while blue circles denote PAH-poor sources (EW7.7\um\,$<$\,0.8\um). \label{ew_hists}}
\end{figure}

The downside is that it is the feature most affected by the continuum model. The 6.2\um\ feature is much cleaner in that respect, and therefore is usually used as proxy for the PAH strength of local LIRGs and \ulig\ \citep[e.g.][]{armus06,desai06}.  In Figure~\ref{ew_hists} we show the restframe 6.2\um\ equivalent width ($EW_{6.2}$) vs. the restframe 7.7\um\ equivalent width ($EW_{7.7}$). We find that $EW_{6.2}$\,$\gs$\,0.2\um\ is the detectability threshold for the 6.2\um\ feature, which translates into $EW_{7.7}$\,$\gs$\,0.8\um. \citet{armus06} find that $EW_{6.2}$\,$\gs$\,0.4\um\ separates pure starbursts from composite systems. Although this is uncertain to $\sim$\,50\% due to the different fitting approach in \citet{armus06}, it suggests that our borderline detected sources are likely to include significant AGN contribution as well.  Therefore, this equivalent width cut should not automatically be translated into a starburst-AGN separation, but merely into a separation into strong-PAH (largely corresponding to detected 6.2\um\ feature when the data are available) and weak-PAH sources.  From now on we refer to sources with $EW_{7.7}$\,$\geq$\,0.8\um\ as {\it PAH-rich}, and those with $EW_{7.7}$\,$<$\,0.8\um\ as {\it PAH-poor}.  We emphasize that PAH-poor does not imply no-PAH since 67\% of the PAH-poor sources also have detectable 7.7\um\ PAH feature, and likely have some star-formation activity in addition to the dominant AGN. A quarter of our sample are PAH-rich sources, and three quarters are PAH-poor sources.  In the rest of this section, we attempt to attach more physical meaning to this empirical classification. In particular, we ask whether, or not, our strong-PAH sources are composite systems with comparable AGN and starburst contribution to the mid-IR light, or some of them are clearly starburst-dominated.  

Lastly, for the purpose of comparison, we have applied our fitting approach on a small sub-sample of the local ULIRG sample from \citet{armus06}. These are overplotted in Figure~\ref{ew_hists}, and will be included whenever possible in subsequent figures. 
\clearpage
\clearpage 
\subsubsection{Trends with \z and luminosity \label{ztrends}}
In PAPERI, we presented the \z distribution of our sample, showing a $z$\,$\sim$\,2.1 peak and a smaller $z$\,$\sim$\,0.8 peak. Given our 24\um\ flux selection, we expect the lower-$z$ galaxies to have \lir{12} (typical local \ulig), while the higher-$z$ galaxies to have up to $L_{\rm{IR}}$\,$\sim$\,\lir{13} (potential HyLIRGs).  Locally the AGN-fraction in local ULIRGs rises with total infrared luminosity, dominating above $\sim$\,3\,$\times$\,\lir{12} \citep{tran01,armus07}. If this holds for our sample, we would expect the strong PAH sources to be among the lower-$z$ sub-sample, while the higher-$z$, higher-$L_{\rm{IR}}$ sources to be continuum-dominated.  
\clearpage
\begin{figure}
\plotone{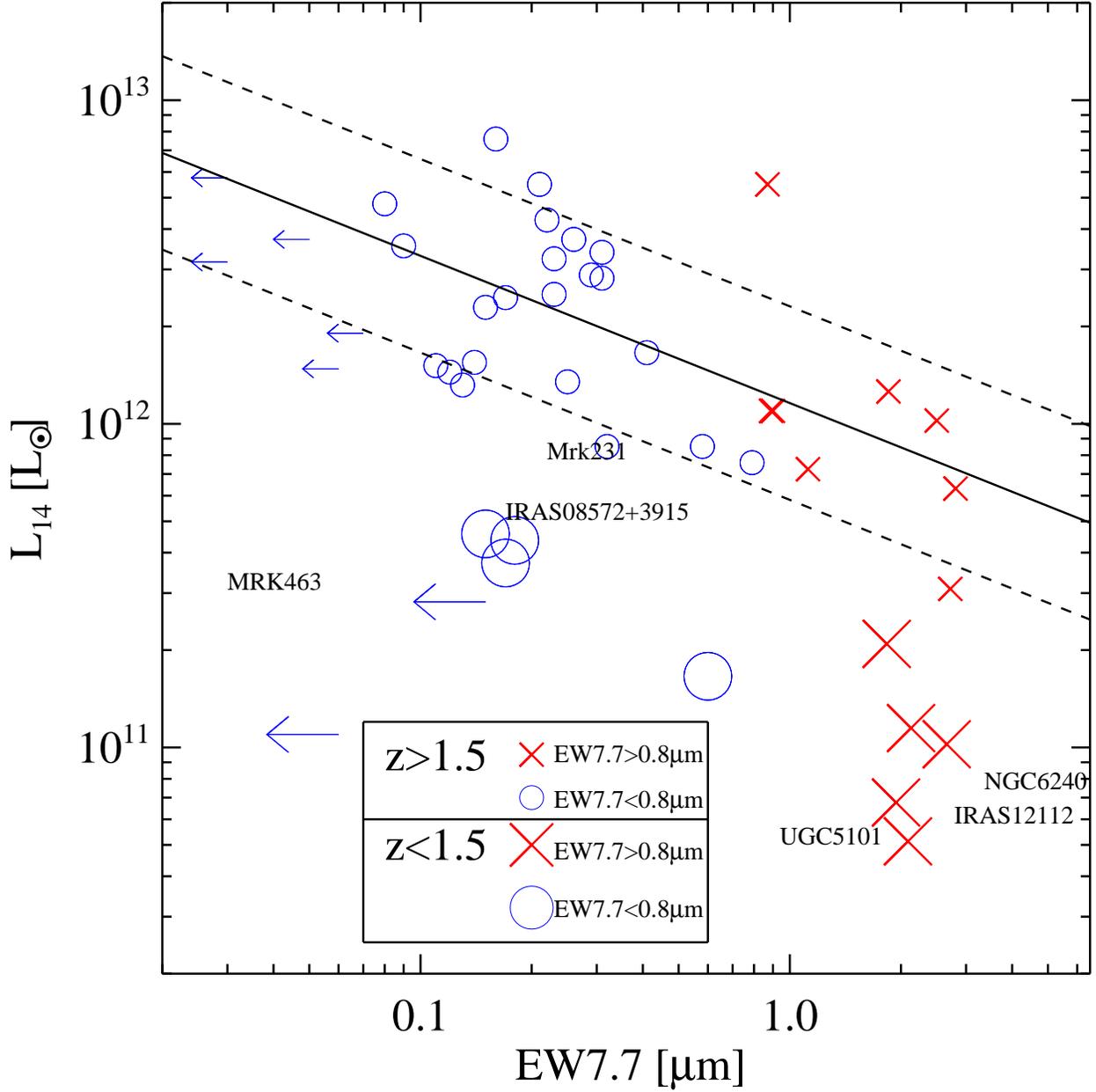}
\caption{For the $z$\,$\geq$\,1.5,  we find a trend of decreasing $EW_{7.7}$ with increasing $L_{14}$. As usual, we mark the strong-PAH sources (i.e. EW7.7\,$\geq$\,0.8\um) with red crosses, while those below this limit are shown with open blue circles. All $z$\,$<$\,1.5 sources (larger symbols) are significantly below the trend found for the higher-$z$ sources. For the whole sample, 25\% are strong PAH sources -- for the low-$z$ ($z$\,$<$\,1.5) sources this fraction is 42\%, while for the high-$z$ sources it is only 23\%. \label{ew7_lum14}}
\end{figure}
\clearpage
Figure~\ref{ew7_lum14}, shows that within a given \z bin, there is a trend of increasing $L_{14}$ with decreasing $EW_{7.7}$.  This might be a trivial result as increasing mid-IR \con by definition decreases the equivalent width. We need total infrared luminosities to study trends in PAH strength with luminosity as seen for local \ulig\ \citep{tran01}. 
\clearpage
\begin{figure}
\begin{center}
\plotone{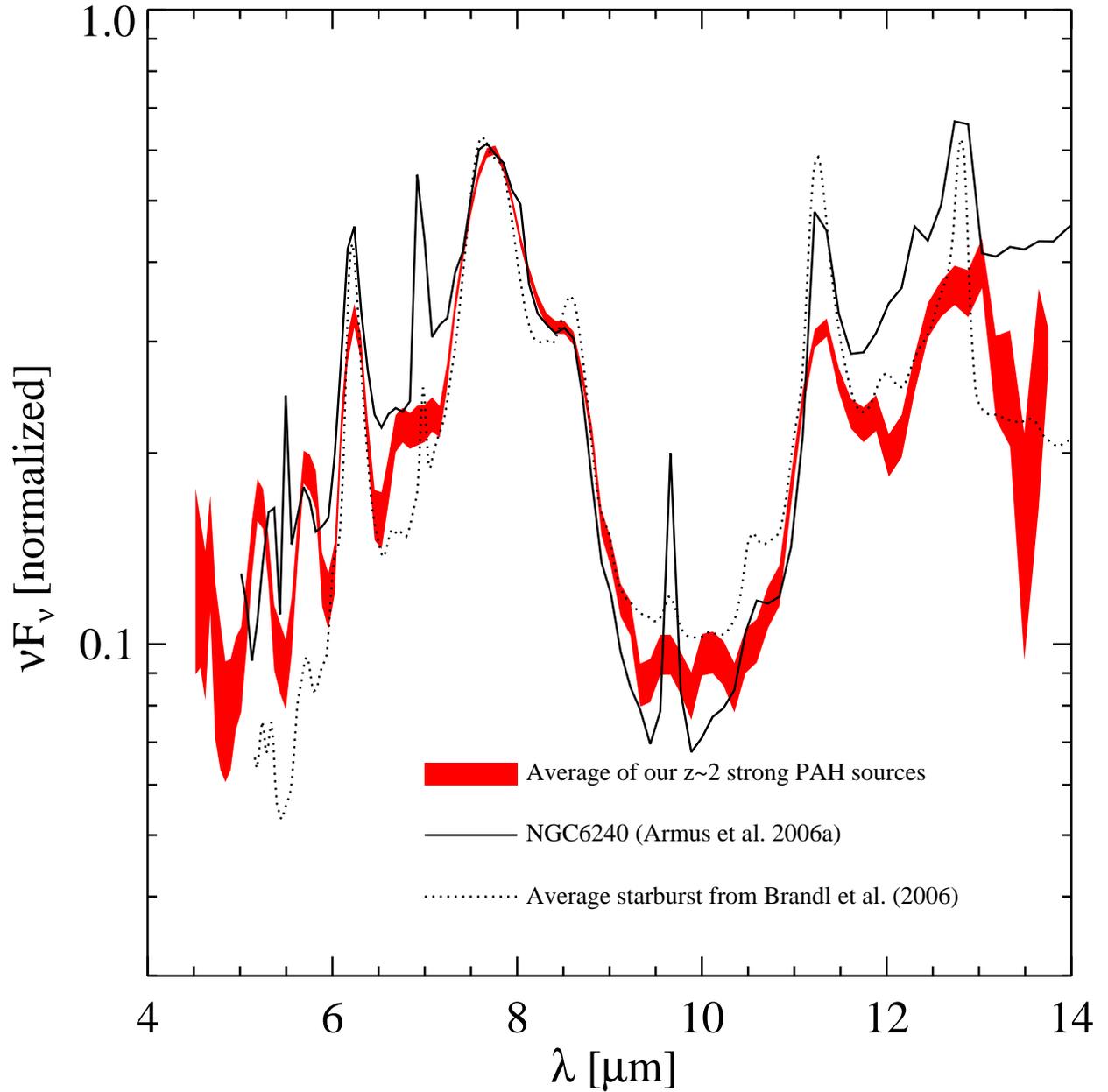}
\end{center}
\caption{The average spectra of four of our $z$\,$\sim$\,2 strong-PAH sources, normalized at 7.7\um.  Our sources appear very similar to the  local ULIRG NGC6240. Both are similar to the starburst template, except for the excess power at $\lambda$\,$<$\,6\um. \label{strongpahstacks}}
\end{figure}
\clearpage
A more intriguing aspect of Figure~\ref{ew7_lum14} is that although it is true that strong PAH sources are predominantly low-$z$, there is a population of equally strong PAH sources at $z$\,$\sim$\,2. Figure~\ref{strongpahstacks} shows the average of strongest-PAH ($EW_{7.7}$\,$\gs$\,2\um) of these $z$\,$\sim$\,2  sources (MIPS289, MIPS8493, MIPS16144, and MIPS22530), normalized at 7.7\um. These sources have $L_{14}$\,$\sim$\,10$^{12}$\lsun, implying  $L_{\rm{IR}}$\,$\sim$\,10$^{13}$\lsun.  For comparison we overlay the average starburst template from \citet{brandl06}, as well as the local ULIRG NGC6240 ($L_{14}$\,$\sim$\,7\,$\times$\,10$^{10}$\lsun, from our own fit) \citep{armus_ngc6240}.  Our $z$\,$\sim$\,2 strong-PAH sources have spectra very similar to NGC6240; however, our sources are about an order of magnitude more luminous.  Lastly, both our average $z$\,$\sim$\,2 spectrum and NGC6240 show excess emission at $\lambda$\,$<$\,6\um\ relative to the average starburst.  We return to this point in \S\,\ref{sec_hotdust}. 

 \subsection{Mid-IR colors: hot and cold dust \label{sec_hotdust}}
The inner dust tori around AGN can be heated to near sublimation temperatures ($\sim$\,1000\,--\,1500\,K), which dominates the mid-IR spectra provided these hot regions are not obscured (e.g. Pier \& Krolik 1992). Such emission peaks in the $\sim$\,3\,--\,6\um\ region leading to (Type 1) AGN to have flatter $\sim$\,5\,--\,15\um\ spectral slopes ($\alpha$\,$\sim$\,1, e.g. 3c273) than the typical starburst ($\alpha$\,$\sim$\,3, e.g. NGC7714). The typical slopes of local ULIRGs \citep{armus06} lie roughly between these values, and are consistent with our median slope of $\alpha$\,$\sim$\,2. We find no difference in the $\sim$\,5\,--\,15\um\ slopes distribution of the PAH-rich and PAH-poor sources.  

\begin{figure}
\plotone{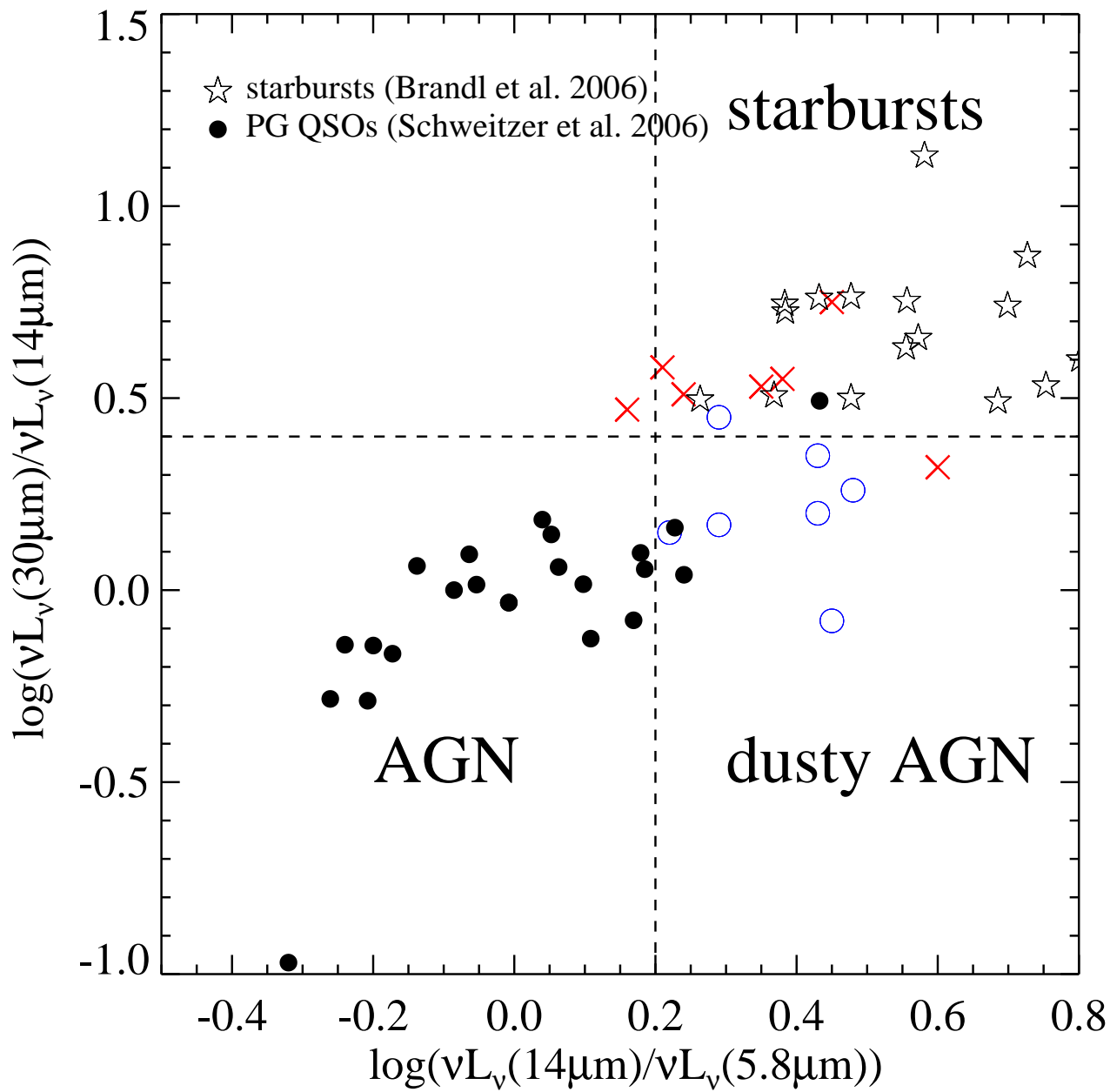}
\caption{ The mid-IR color-color diagram, where only $z$\,$<$1.6 sources are shown as  reliable 30\um\ luminosities cannot be obtained for the rest. For comparison we overlay local samples of starbursts and optically-selected quasars. The dashed line is to guide the eye.  Note that among the reddest in both colors sources, the two populations are indistinguishable.  \label{luminos}}
\end{figure}

The longer wavelength SED is dominated by cold dust  of $\ls$\,50\,K associated with star-formation.  In \S\,\ref{sec_excess} and Figure~\ref{mirspec}, we show that many of our sources show very warm spectra peaking at $\sim$\,30\um, while others show an excess over the mid-IR continuum, consistent with the presence of cooler dust. These were used to derive $\nu L_{\nu}$(30\um) (see Table~\ref{lums_table}) for sources with $z$\,$\ls$\,1.6 such that the MIPS70\um\ point is not too redshifted away from restframe 30\um.  Figure~\ref{luminos}, shows that while the $\sim$\,5\um-to-15\um\ colors are the same for both PAH-rich and PAH-poor sources (as is the case for the entire sample), the $\sim$\,15\um-to-30\um\ colors separate the two groups very well. Our PAH-poor sources are redder than the optically-selected quasars consistent with their being dustier. Our PAH-rich sources are generally consistent with the starbursts' colors although they show slightly bluer $\sim$\,5\,--\,15\um\ slopes suggesting enhanced hot dust. 
\clearpage

\begin{figure}
\plotone{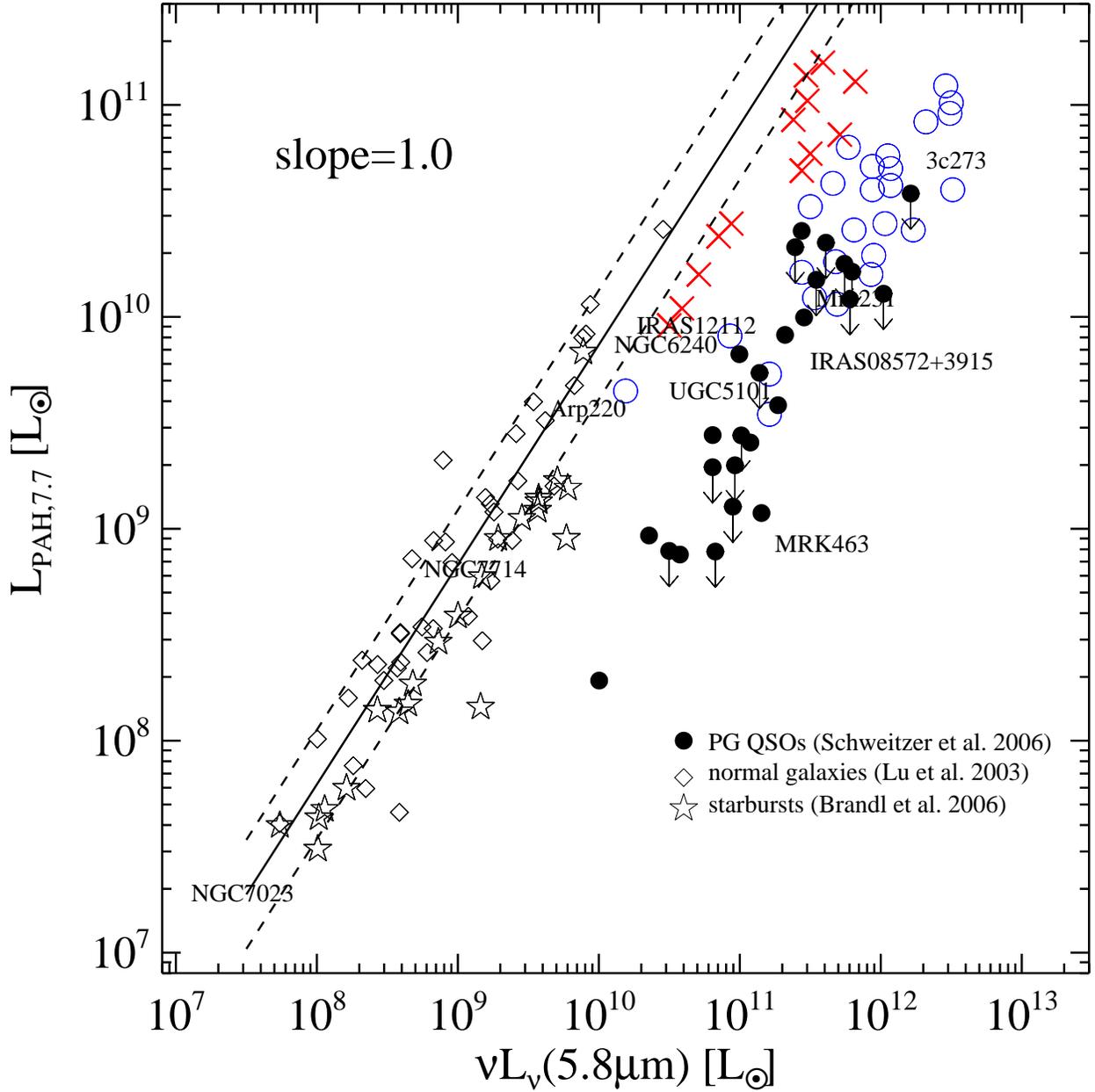}
\caption{ The $\nu L_{\nu}(5.8)$ vs. $L_{\rm{PAH},7.7}$. For comparison we overlay local samples of optically-selected AGN, normal star-forming galaxies, and starbursts. The exact positions of these are uncertain as we have converted the published PAH 7.7\um\ feature strength values to our continuum definition. We have fitted the named sources with our method. \label{lum5_lum7}}
\end{figure}
\clearpage
We expand on the last point,  now using the entire sample, in Figure~\ref{lum5_lum7}, where we show the hot dust continuum luminosity $\nu L_{\nu}(5.8)$ vs. the 7.7\um\ PAH luminosity. For lower luminosity star-forming sources, the two are well correlated with slope unity suggesting at least {\it at moderate luminosities}, the mid-IR SEDs of purely star-forming galaxies are fairly constant. As expected, optically-selected quasars have stronger 5.8\um\ \con per 7.7\um\ luminosity. Our weak-PAH sources, at higher luminosities, are consistent with the local AGNs trend, despite their redder slopes. Although, to a lesser extend, our strong-PAH sources  are also noticeably shifted with respect to the local starbursts. There are two possible explanations for this:
 
1) They have significant AGN contributions leading to enhanced 5.8\um\ continuum. 

2) The local starburst trend does not extend to the highest luminosities regime, but curves toward our strong-PAH sources making them starburst-dominated as well. In other words, extreme starbursts generate enhanced hot dust emission compared with lower luminosity ones. 

The first explanation is very likely given the extreme luminosities of our sources. As discussed in \S\,\ref{ew7}, our equivalent width cut is more an upper limit on the starburst-dominated sources, since especially borderline sources are likely to include significant AGN contribution as well. 
Therefore, we also look individually at the $z$\,$\sim$\,2 sources with strongest PAH emission ($EW_{7.7}$\,$\gs$\,2\um, or $EW_{6.2}$\,$\gs$\,0.3\um).  Even these strongest PAH sources, show excess emission at $\lambda$\,$<$\,6\um, relative to the typical starburst (see Figure~\ref{strongpahstacks}), indicating higher hot dust contribution.  Figure~\ref{strongpahstacks} also shows that our strong-PAH sources are similar to NGC6240 in that respect, which was already interpreted by \citet{armus_ngc6240} as having $\sim$\,20\% AGN contribution to the bolometric luminosity ($\sim$\,50\% of the mid-IR emission; \citet{lutz03}). 

On the other hand, hot dust emission has been observed to be associated with star-formation activity as well \citep{hunt02,lu03}.  An extreme example, are high SFR intensity dwarf galaxies which show continuum-dominated mid-IR spectra \citep{galliano03}. Therefore, the possibility of enhanced hot dust emission being associated with extreme starbursts cannot be excluded. 

We conservatively adopt the first explanation above, but stress that the properties of high-luminosity, pure-starbursts remain unclear.     

\subsection{Near-IR host galaxy light \label{sec_host}}
It has been proposed that the presence of the 1.6\um\ bump, characteristic of stellar populations older than $\sim$\,100\,Myrs, is an efficient way of separating starburst- from AGN-dominated sources \citep[e.g.][]{weedman06,teplitz07}. The rationale is that the hot dust \con of AGN overwhelmes their host galaxy's stellar bump. 
\clearpage
\begin{figure}
\plottwo{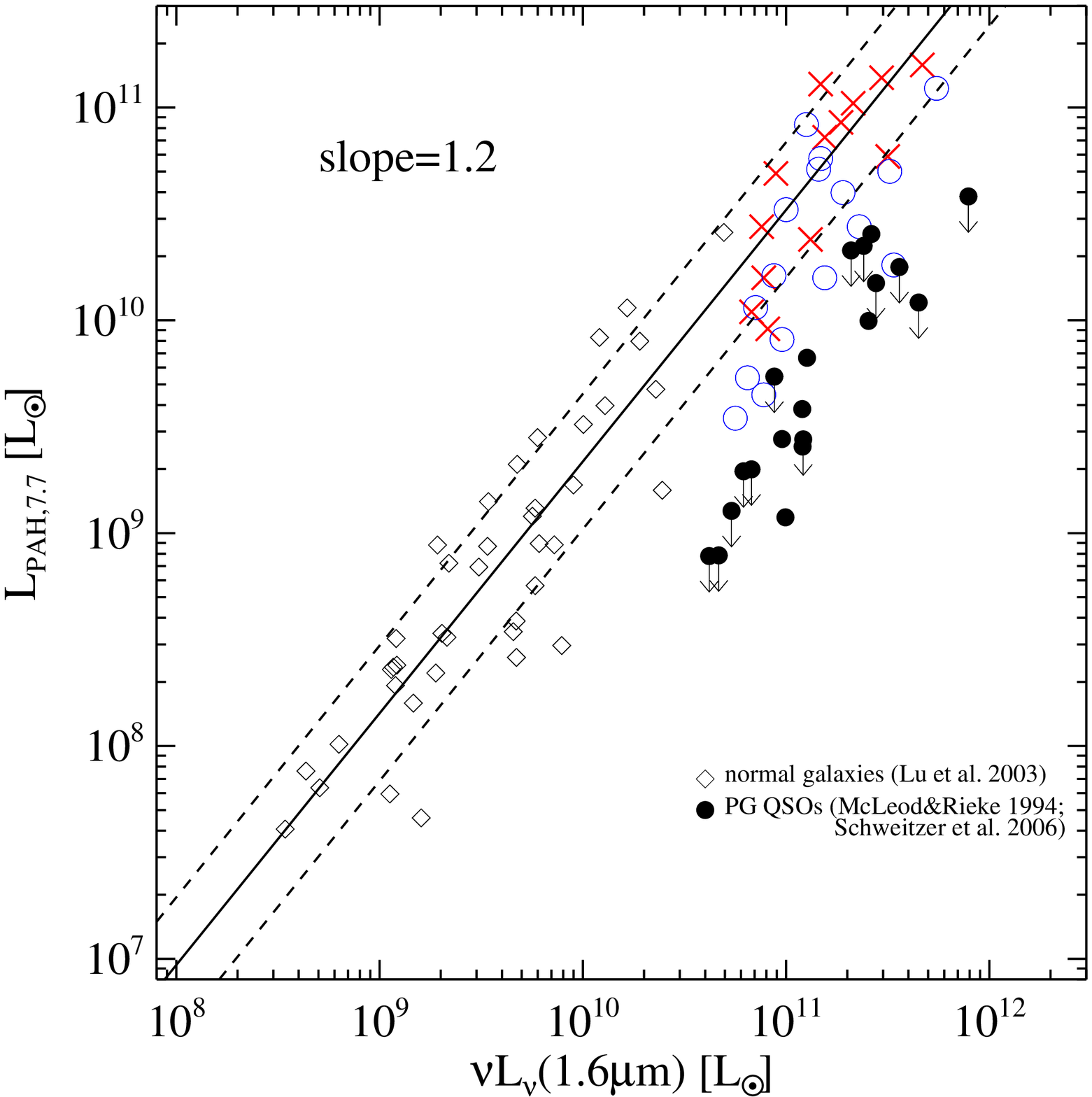}{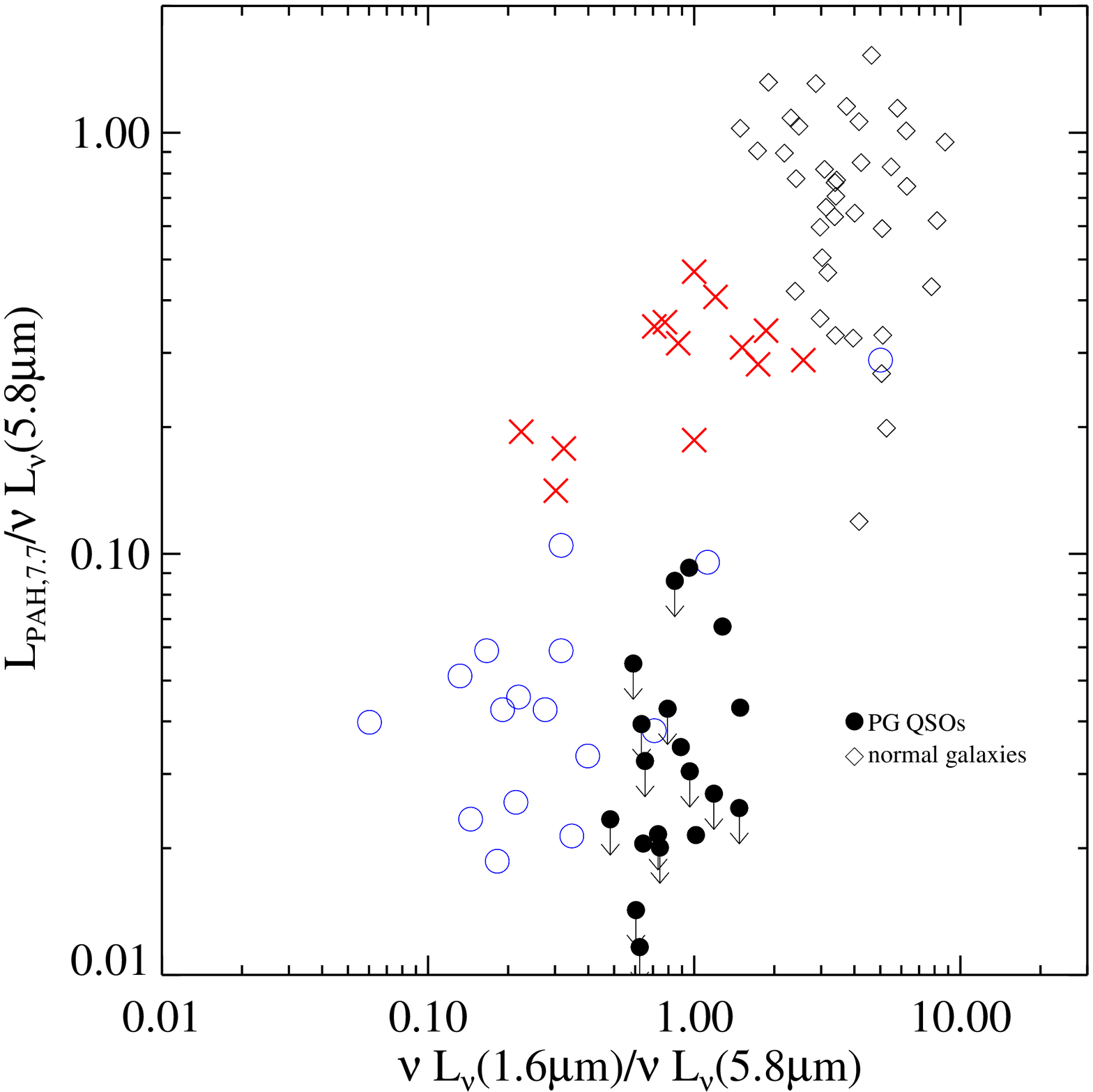}
\caption{ {\it Left:} The 1.6\um\ stellar bump strength vs. the 7.7\um\ PAH strength. The normal starforming galaxies are used to define the best-fit line shown. Note that both our PAH-rich and PAH-poor sources agree very well with this relation.  Optically-selected AGN light to the right of this, suggesting that for the bulk of our PAH-poor sources, the 1.6\um\ emission is more likely to be due to host galaxy emission rather than that of the AGN itself as in the PG QSOs sample.  {\it Right:} The 1.6\um\ stellar bump strength vs. the 7.7\um\ PAH strength normalized to the 5.8\um\ luminosities. \label{lumh_lum7}}
\end{figure}
\clearpage
Figure~\ref{mirspec} shows that the IRAC points (which probe the 1.6\um stellar bump at $z$\,$\sim$\,2) often (in 73\% of the sources) have excess emission over the extrapolated mid-IR continuum.  This excess is most likely the 1.6\um\ stellar bump of old stellar populations together with some unknown young stars contribution (see \S\,\ref{sec_excess}).  The left-hand panel of Figure~\ref{lumh_lum7} shows the 1.6\um\ luminosity vs. the PAH luminosity. Both our PAH-rich and PAH-poor sources (where the excess is observed) obey the relation defined by the local star-forming galaxies.  This suggests that even in the continuum-dominated sources (with a few exceptions) we do not see the AGN itself (accretion disk emission), but the host galaxy light alone\footnote{Upcoming NICMOS observations of the sample would be able to address this directly}. Circumstantial support is provided by optically-selected quasars whose 1.6\um\ luminosities (McLeod \& Rieke 1994a,b) lie significantly to the right of the star-forming galaxies relation. Although not shown for clarity, the AGN-subtracted H-band luminosities of the PG QSOs (McLeod \& Rieke 1994a,b) agree much better with the star-forming galaxies relation suggesting the offset is predominantly due to the effect of AGN-dominance at 1.6\um\ rather than suppressed 7.7\um\ emission. 

The last point is supported by the right-hand panel of Figure~\ref{lumh_lum7} shows $\nu L_{\nu}$(1.6\um) vs. the 7.7\um\ PAH feature luminosity both normalized to the 5.8\um\ continuum. The effect of increasing hot dust emission is now obvious, as there is a clear trend between the starforming galaxies, our PAH-rich and PAH-poor sources. The PG QSO sample have the same 7.7\um\ feature strength relative to the 5.8\um\ as our PAH-poor sources, but show an excess of 1.6\um\ emission.  
\clearpage
\begin{figure}
\plotone{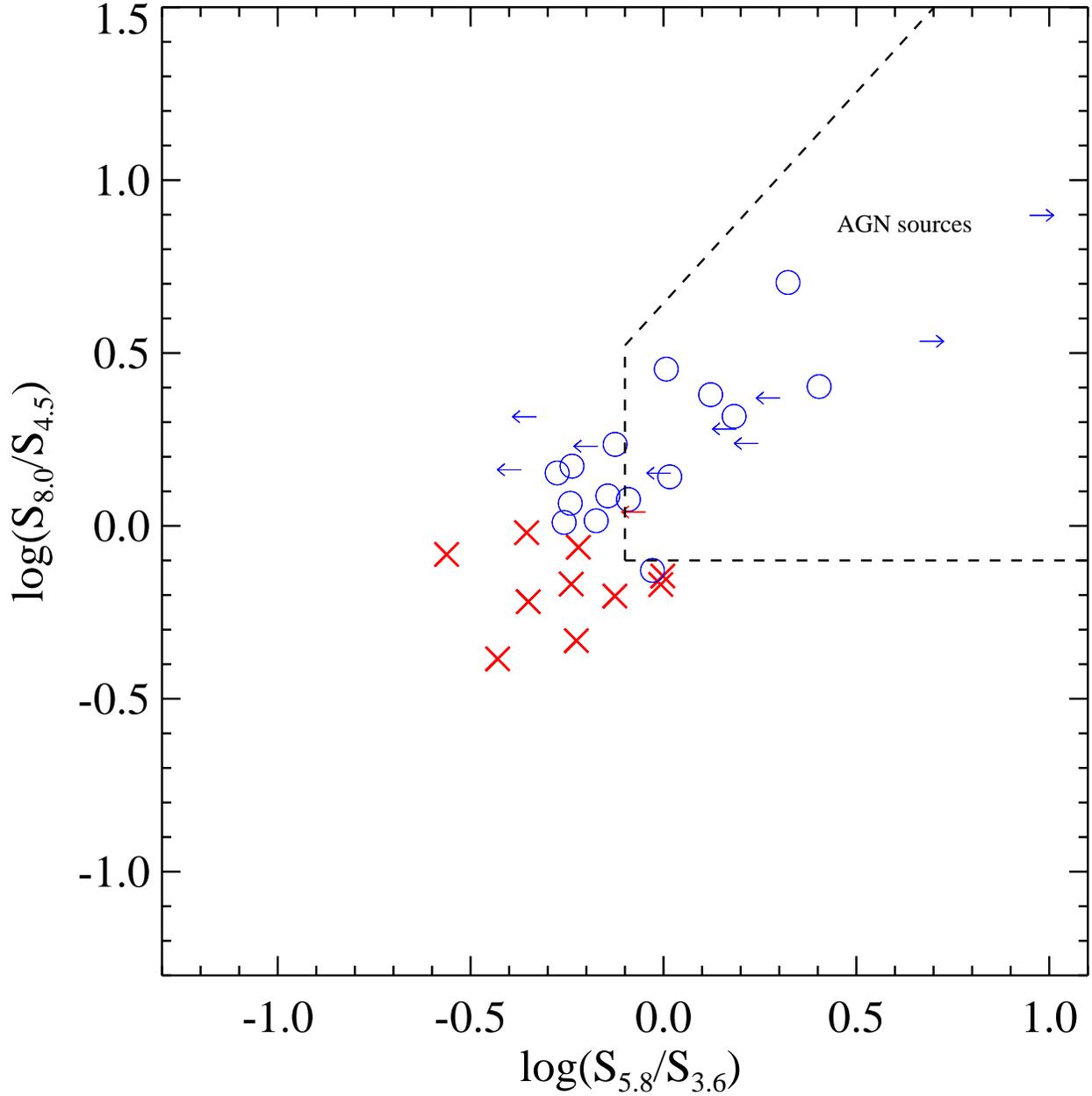}
\caption{ The IRAC color-color plot of our sample, excluding sources with more than one band at $<$\,3\,$\sigma$.  The colors as usual mean PAH-rich (red crosses) and PAH-poor (blue circles). The dashed lines indicate the wedge used in selecting AGN-dominated sources at lower-$z$ \citep{lacy04}.  \label{irac_ccd}}
\end{figure}
\clearpage
Lastly, these results have implications for AGN-selection methods such as the IRAC color-color diagram (see Lacy et al. 2004, Stern et al. 2005, Sajina, Lacy, Scott 2005, Lacy et al. 2006). These rely on the power-law continuum in the IRAC band of AGN, and the PAH, and stellar bump features in the spectra of starbursts. In Sajina, Lacy, \& Scott (2005), we stressed that this might fail at $z$\,$\sim$\,2 due to the unknown fraction of AGN whose host galaxies dominate at $\lambda$\,=\,1.6\um.  Figure~\ref{irac_ccd} shows the IRAC color-color plot for our sample.  About 20\% of the continuum-dominated sources fall outside of the AGN regime due to host galaxy emission as discussed above.  This assumes that the sources undetected in more than 1 IRAC band and therefore not included in Figure~\ref{irac_ccd} are undetected due to steeply falling red continua and would have fallen inside the AGN-wedge.  Since our sample is biased toward less prominent hot dust, this is an upper limit for the AGN population as a whole. It does however, imply that the use of this technique to search for obscured AGN, although remarkably effective at lower-$z$ \citep[e.g.][]{lacy06} becomes less reliable at $z$\,$\gs$\,2.

\subsection{The 9.7\um\ silicate feature}
The depth of the 9.7\um\ silicate feature has recently been highlighted \citep{levenson,newdiag,imanishi} as a tool to distinguish between sources dominated by buried nuclei from sources with more clumpy, extended dust distributions.  \citet{newdiag} presented an intriguing diagnostic diagram of the PAH6.2 equivalent width vs. the 9.7\um\ silicate feature depth. Sources with weak silicate features, range from pure starbursts to pure AGN in the manner of traditional PAH-over-continuum diagnostics. A distinct secondary sequence of increasing silicate feature depth with decreasing PAH feature strength is revealed.  This is interpreted as an increasingly dominant buried nucleus.
 
\begin{figure}
\plotone{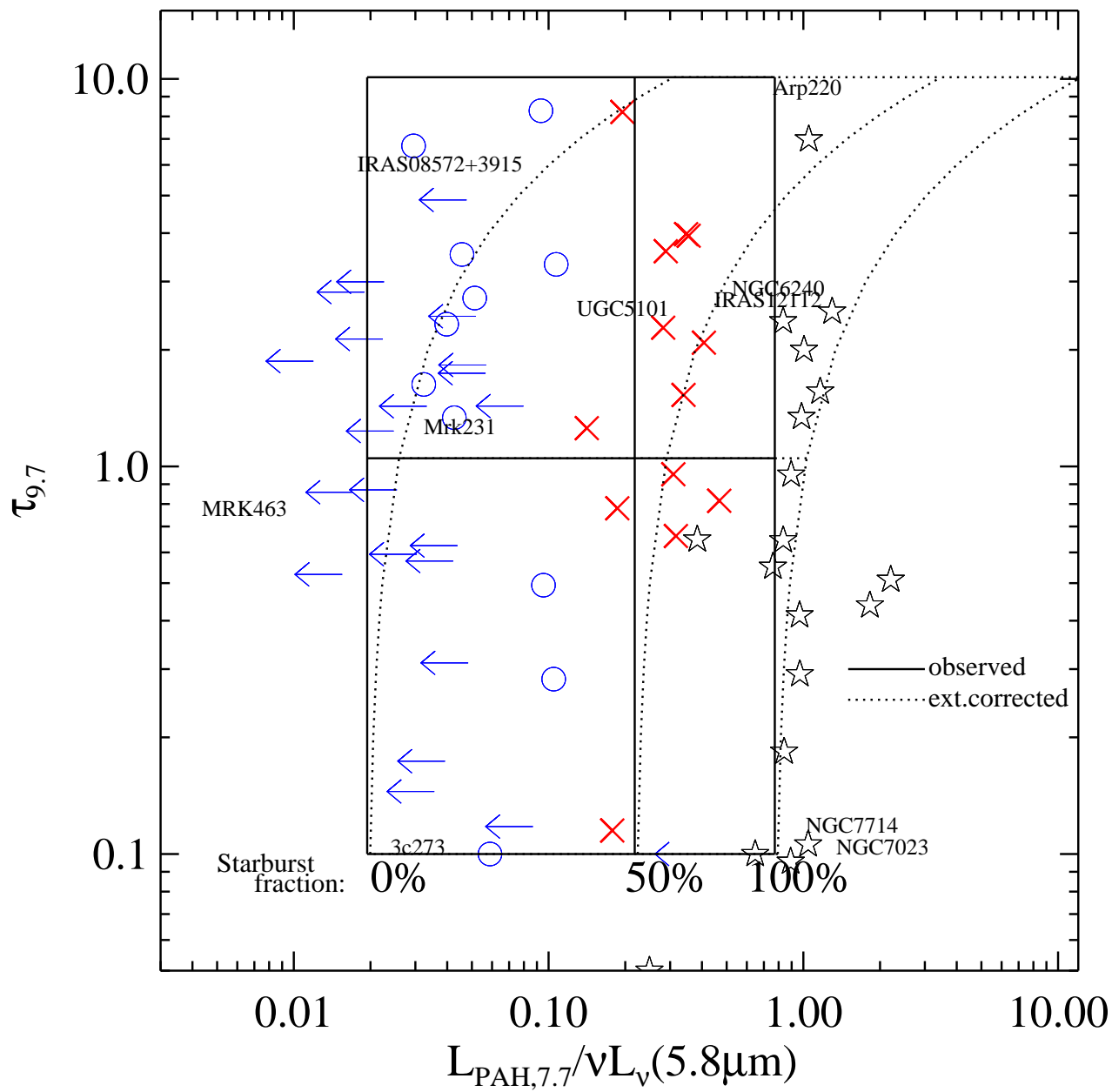}
\caption{The low-$\tau$, x-axis starburst fraction span the range from pure starburst (NGC7714) to pure typeI AGN (3c273). Maintaining the same {\it observed} fraction with increasing $\tau$ gives us the solid lines. Maintaining the same {\it intrinsic} fraction with increasing $\tau$ gives the dotted lines.  The stars are the Brandl et al. (2006) starbursts. \label{pah_tau}}
\end{figure}

 For low signal-to-noise, high-$z$ samples such as ours, the equivalent diagram is based on the 7.7\um\ PAH feature strength, and the optical depth estimate (Figure~\ref{pah_tau}).  The overlaid grids are generated via Equation~1, with an $\alpha$\,=\,2 continuum, and starburst fractions spanning between 3c273 (0\%) and NGC7714 (100\%).  As we increase the optical depth the observed fractions remain constant, while the intrinsic fractions curve since the un-extincted $L_{5.8}$ is higher than the observed one.  We mark $\tau_{9.7}$\,=\,1 as separating optically thin from optically thick sources, which is also where extinction correction effects become strong. This largely corresponds to the separation between class 1 and class 2 sources in \citet{newdiag}, or in other words between sources with clumpy dust (weak silicate feature) or buried nucleus sources (significant silicate feature). 
\clearpage
\clearpage 
Roughly 40\% of the sample have $\tau_{9.7}$\,$<$\,1, indicating that a buried nucleus is not present or significant.  Of these sources, three quarters are continuum-dominated (most likely AGN) sources.  A quarter are strong PAH sources, but none of them are likely to be pure starbursts and most likely have significant AGN contribution as well. 

The bulk of our sample, $\sim$\,60\%, have $\tau_{9.7}$\,$>$\,1, suggesting that the buried nucleus dominates over more clumpy dust geometries.  Although about 25\% of these sources have significant PAH emission, after extinction correction $\sim$\,90\% of these highly obscured sources are continuum-dominated. It is important to realize that since our sensitivity to weak PAH emission (even using the 7.7\um\ feature) is worse than for the local sample used by Spoon et al., our diagram is missing (or are lost in the upper limits) the weakest PAH sources (class A in Spoon et al.).  The 50\% line roughly corresponds to the separation between class B and class C sources in Spoon et al.  

Lastly, we compared the relative distribution of our sources along the $\tau$-direction with that of local ULIRGs \citep{newdiag}. We find that, the two are roughly consistent with both populations having the bulk of their members residing in the moderate optical depth regime ($\tau_{9.7}$\,$\sim$\,1\,--\,3).  For both local and high-$z$ ULIRGs (including those with only lower limits on $\tau_{9.7}$), $\sim$\,25\% of the population resides in the most obscured sources regime ($\tau_{9.7}$\,$\gs$\,3). This corresponds to class 3 in \citet{newdiag}, which contains the sources with evidence of extremely cold and dense media (see Appendix~A).  We estimate that for the local sample, the space density of these extremely obscured sources to be roughly $\sim$\,6\,$\times$\,$10^{-8}$\,Mpc$^{-3}$ (assuming that the ULIRG sample in Spoon et al. (2006) is not biased in its $\tau$-distribution compared with the IRAS 1Jy ULIRG sample).  Given that our sample of $z$\,$\sim$\,2 extremely obscured sources is incomplete, we derive a lower limit density of $\sim$\,7\,$\times$\,$10^{-7}$\,Mpc$^{-3}$. These estimates are uncertain to within a factor of 2, but clearly suggest that the density of such objects was about an order of magnitude larger at $z$\,$\sim$\,2 than today. 

\section{Summary \& Conclusions}

We presented spectral decomposition of a sample of 48 $z\sim2$, low resolution mid-IR IRS spectra. The detailed analysis, separating the contribution of PAH emission, \con emission and \con extinction, has revealed several significant characteristics of $z\sim2$ \ulig.  

In PAPERI, we discussed that our sources have quasar-like luminosities, and frequently continuum-dominanted spectra suggesting an AGN is the primary power source. We find that 75\% of our sample have continuum-dominated spectra. However, four sources of the original sample of 52 were not included here because of their featureless spectra and hence no redshifts. If we include these 4 sources, this fraction rises to nearly 80\%. About 60\% of these show detectable PAH emission, suggesting these are composite sources rather than pure AGN. At much higher signal-to-noise ratio, \citet{schweitzer06} find $\sim$\,40\% of PG QSOs have detectable PAH. We have shown that our sources have much redder mid-IR colors than the PG quasar sample. Lastly, in optically-selected quasars, the AGN dominates in the near-IR, whereas for us the host galaxy, or a significantly obscured AGN,  appear to dominate at $\sim$\,1\,--\,2\um.  The greater obscuration is also seen in the fact that $\sim$\,60\% of these continuum-dominated sources have $\tau_{9.7}$\,$>$\,1 (see discussion below), whereas optically-selected quasars have $\tau_{9.7}$\,$\sim$\,0 (and even show silicate emission). All of these suggest that the bulk of our sources are potentially related to the still elusive type-2 quasars. Follow-up observations are needed to ascertain their nature.    

About 25\% of our sample  have strong-PAH. In particular, for 38 spectra covering the rest-frame 6\um\, six sources have detectable 6.2\um\ PAH emission with $EW_{6.2}$ ranging $\sim$\,0.3\,--\,0.5\um.  Their 14\um\ monochromatic luminosities are on the order of $10^{12}L_\odot$.  \ulig\ with such a high infrared luminosity and such a large 6.2\um\ PAH equivalent width do not exist in the local Universe. As shown by the IRS observations of local \ulig\ \citep{armus06},  a galaxy with 6.2\um\ PAH equivalent width
of 0.3\um\ has 14\um\ luminosity $\sim10^{11}L_\odot$, an order of magnitude less than that of our $z\sim2$ PAH emitting sources.  Regardless of the exact degree of AGN contribution, it is obvious that the local PAH6.2\um\ vs. $L_{\rm{IR}}$ relation (Armus et al. 2007 in prep.) does not hold at $z$\,$\sim$\,2 (or experiences strong evolution).  These $z\sim2$ strong-PAH sources likely all have significant AGN contribution to their mid-IR emission, as evidenced by excess hot dust emission. However, their bolometric emission is likely still dominated by the starburst (by analogy with sources such as NGC6240, and the observed hints of colder dust at longer wavelengths). Far-IR observations are needed to disentangle the star-formation and AGN contribution to their infrared luminosity (see Sajina et al. 2007 in prep.).  If these sources are indeed starburst-dominated, translating their \con luminosities into bolometric luminosities (see PAPERI) implies SFR on the order of 1000$M_{\odot}/yr$ (this could be lowered significantly by including an obscured AGN, or top-heavy IMF \citep{baugh05}). Such extreme values have been claimed before for sub-mm galaxies \citep[SMGs; see e.g.][]{blain02,pope06}, which have comparable number density at $z$\,$\sim$\,2 as our sources (PAPERI).  

The bulk of these luminous sources show significant obscuration with optical depths $\tau_{9.7}$\,$>$\,1. We propose a new diagnostic diagram, applicable to low signal-to-noise ratio high-$z$ data, where we plot the optical depth against the traditional PAH strength indicator of  relative starburst/AGN activity. Following Levenson et al. 2006 and \citet{newdiag},  about 60\% of our sample are consistent with being dominated by  buried nuclei rather than the clumpy dust distribution indicating (but not exclusively) extended star-formation activity. About 50\% of our strong-PAH sources fall in this category.  However, we find some $z$\,$\sim$\,2, $L_{\rm{IR}}$\,$\sim$\,$10^{13}$\lsun\ strong-PAH sources that are potentially consistent with extended, clumpy star-formation (e.g. MIPS289). Recent observations suggest this may be the case with some SMGs as well \citep{ivison07}.  The definitive answer to this question will have to await the greater spatial resolution in the sub-mm regime of the Atacama Large Millimeter Array \citep[ALMA;][]{alma}. 

Among our sources, the distribution of the absorption depth at 9.7\um\ is similar to what observed among local \ulig, with $\langle$\,$\tau_{9.7}$\,$\rangle$\,$\sim$\,1.5; however, our sources on average are a factor of 5-10 more luminous ($L_{8-1000\mu m}$) than local \ulig, and close to the bolometric luminosity of the brightest, optically selected quasars.  The relative starburst-AGN contribution to the mid-IR emission of these sources is difficult to assess, especially for the most extremely obscured sources ($\tau_{9.7}$\,$\sim$\,3), which represent $\sim$\,25\% of our sample. The strong similarity between the mid-IR spectrum of the nascent starburst NGC1377 \citep{roussel06} and the deep silicate feature ULIRG IRAS08572+3915 suggest  that even in sources with weak or absent PAH, a buried nucleus can be powered by either AGN or a young starburst (or more likely both). High resolution mid-IR spectra with Neon line ratios, deep X-ray spectra, L-band (3\,--\,4\um) spectroscopic diagnostics could provide useful additional information, but uncertainties due to high obscuration would remain \citep[see e.g. discussion on Arp220,][]{spoon_arp220}.  If this deeply obscured phase is one of the ULIRG/quasar evolutionary stages, this phase is unlikely to last longer than $\sim$\,$10^6$\,--\,$10^7$yrs, or until the supernovae- or AGN-driven winds punch holes through the cacoon.  Establishing the frequency of such sources, compared with the clearly starburst-like and AGN-like ULIRGs, any possible evolution with redshift and luminosity would be the first steps to test whether this is a necessary evolutionary stage that all ULIRGs pass, or an alternate pathway depending on the properties of the merging components. From our sample, we estimate that $z$\,$\sim$\,2 extremely obscured sources, have a space density of at least $\sim$\,7\,$\times$\,$10^{-7}$\,Mpc$^{-3}$\,ster$^{-1}$ or about an order of magnitude grater than locally. 

Our study hints at some of the specific physical processes behind the strong evolution of the infrared luminosity function between $z$\,$\sim$\,2 and today \citep{lefloch05,caputi07}.  Although our biased selection precludes a robust comparison, we find that $z$\,$\sim$\,2 ULIRGs have roughly consistent spectral properties (PAH strength, mid-IR slope, and SiO feature depth) as local ULIRGs, but are an order of magnitude more luminous. Thus the implied luminosity evolution does not follow local trends with luminosity, which would for example preclude the existence of strong-PAH sources with $L_{\rm{IR}}$\,$\sim$\,$10^{13}$\lsun.  The fraction of extremely obscured sources remains constant, but their space density increases by about an order of magnitude suggesting that number evolution plays a role as well as luminosity evolution.  Lastly, whether our sources are scaled-up local ULIRGs, or driven by fundamentally different modes of star-formation and/or black hole accretion remains an open question. 

\acknowledgements
We would like to thank Vassilis Charmandaris, Guilaine Lagache and Daniel Stern for useful discussions. Many thanks to Kalliopi Dasyra for help with PG QSO data and useful discussions.  We would like to thank the anonymous referee for helpful suggestions which have improved the presentation of this paper. 
This work is based on observations made with the {\sl Spitzer} Space Telescope, which is operated by the Jet Propulsion Laboratory, California Institute of Technology under contract with NASA. Support for this work was provided by NASA through an award issued by JPL/Caltech. A. Sajina acknowledges support by NASA through grant 09865 from the Space Telescope Science Institute, which is operated by the Association of Universities for Research in Astronomy, Inc, under NASA contract BAS5-26555.

\appendix
\clearpage
\begin{figure}
\begin{center}
\plotone{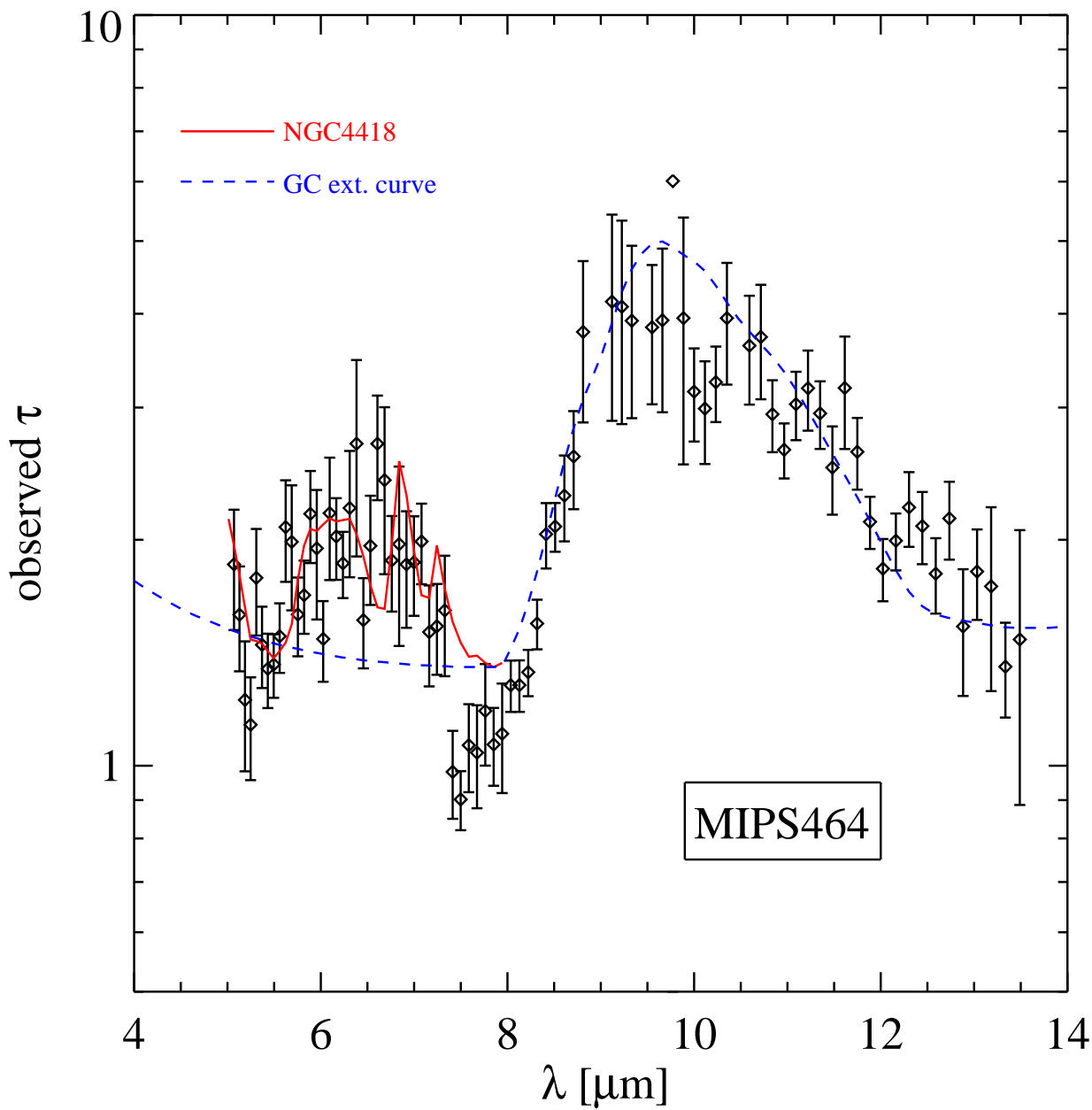}
\end{center}
\caption{The absorption profile of MIPS464 after removing the best-fit continuum.  We overlay the Galactic Center (GC) extinction curve. There is noticeable excess at $\sim$\,6\,--\,7\um. This is most likely due to water ice and HAC absorption features. For comparison we overlay the 5\,--\,8\um\ optical depth profile of NGC4418 scaled to the 9.7\um\ depth of MIPS464 (which is $\sim$\,85\% that of NGC4418).  Despite the low signal-to-noise ratio of our data,  the agreement between the two suggests significant ice absorption in MIPS464.  The dip at $\lambda$\,$\sim$\,8 is due to potential weak PAH component in MIPS464. The dip at $\lambda$\,$\sim$\,10\um\ is likely due to H$_{2}$ S(3) emission.  \label{ice_profile} }
\end{figure}   
\clearpage
\section{Water ice and crystalline silicate features in $z$\,$\sim$\,2 ULIRGs \label{si_shape}}
About 25\% of our sources show extreme silicate absorption ($\tau$\,$\gs$\,3) consistent with the deeply obscured sources in \citet{spoon02}, or class 3A, 3B in \citet{newdiag}.  The high-resolution data available for the local sources discussed in Spoon et al. (2006) show that this class of sources frequently show additional absorption features  consistent with extremely cold, dense media. These include in particular ices and crystalline silicates. Here we discuss evidence that these might be present in the most deeply obscured of our $z$\,$\sim$\,2 sources.  

As mentioned in \S\,\ref{cont-fit},  a number of our sources show $\sim$\,6\,--\,7\um\ absorption in addition to the extinction curve. Most of these are within the 1\,$\sigma$ uncertainty of the data. The source in which this is most prominent is the highly-obscured MIPS464 ($\tau_{9.7}$\,$\sim$\,4.9, $z$\,=\,1.85).  Figure~\ref{ice_profile} shows the  optical depth profile of MIPS464 compared with the Galactic Center extinction curve. In addition, we overlay the 5\,--\,8\um\ optical depth profile of NGC4418 showing the broad 6.2\um\ water ice absorption feature as well as the narrower HAC absorption features at 6.9\um\ and 7.3\um.  The NGC4418 profile has been scaled down by 15\% to match the MIPS464 optical depth at 9.7\um.  Although the signal-to-noise of our data does not allow us the detailed feature fitting that can be done for NGC4418, the good agreement between the two optical depth profiles suggests that MIPS464 is consistent with having water ice absorption as well. Recent studies show that water ice absorption is observed in most local ULIRGs with high silicate feature optical depths ($\tau_{9.7}$\,$\gs$\,3) \citep{imanishi}.  The inferred 6.0\um\ absorption depth of MIPS464 (see Figure\,\ref{ice_profile}) is about twice the average of the \citet{imanishi} sample meaning that this feature is likely to be at $\ls$\,1\,$\sigma$ for most of our highly obscured sources.  
 
\begin{figure}
\begin{center}
\plotone{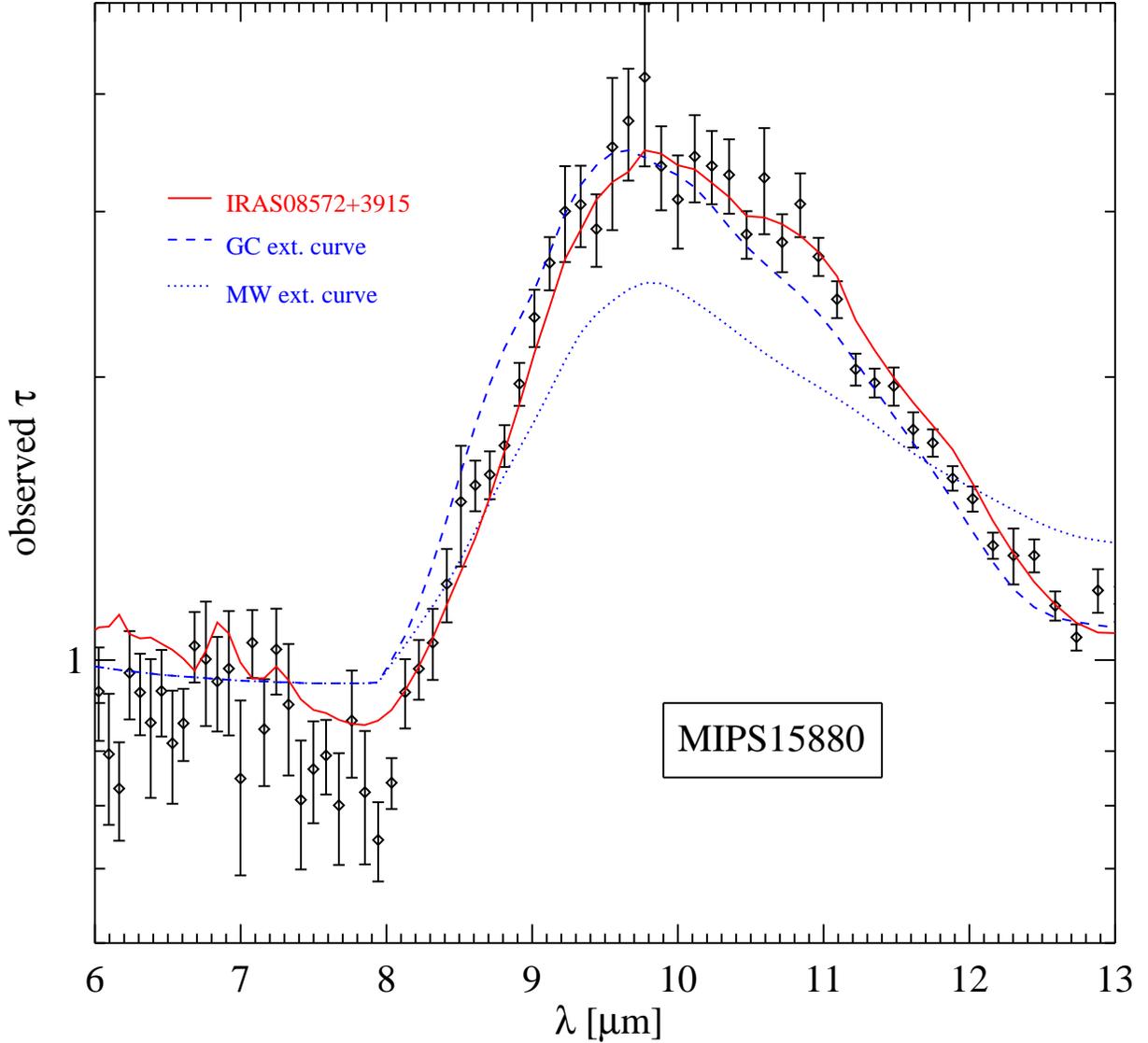}
\end{center}
\caption{The silicate feature profile of MIPS15880 after removing the best-fit continuum. Extinction curves at best only approximate the observed profile. The IRAS08572 spectrum with {\it its} best-fit \con removed, but scaled to the MIPS15880 optical depth. The two show very similar Si feature profiles. The discrepancy at $\lambda$\,$\ls$\,8 is due to weak PAH component in MIPS15880, whereas IRAS08572 does not require any PAH.  The PAH have not been removed from the MIPS15880 profile in order to highlight the excess opacity at 11\um\ even {\it before} the 11.3\um\ PAH feature is removed. \label{si_profile} }
\end{figure}

We also examined our highest signal-to-noise spectra for the presence of crystalline substructure on the otherwise smooth 10\um\ amorphous silicate feature.  In the general ISM of our galaxy the timescale for transformation of crystalline silicates into amorphous silicates is estimated to be short compared w ith the timescale for injection. This is in agreement with the strict upper limit of $<$\,0.01 for the crystalline fraction of the ISM in the line of sight to our Galactic Center \citep{kemper04}.  Recently, \cite{cryst_si} found the first evidence for crystalline silicates in 12 local ULIRGs. Surprisingly, they found a crystalline fraction of $\sim$\,0.1, or about an order of magnitude higher than in the Milky Way disk. Figure~\ref{si_profile} shows the silicate feature profile of MIPS15880 ($\tau_{9.7}$\,$\sim$\,3.9, $z$\,=\,1.64), which has the highest SNR of our sample. The observed profile is remarkably similar to that of IRAS08572+3915, including excess absorption at 11\um\ which for IRAS08572+3915 has been identified with a crystalline silicate feature.  Given the errorbars, this opacity bump is only $\sim$\,2\,$\sigma$.  The 11.3\um\ PAH feature has not been removed, in order to emphasize the significance of this potential excess opacity.  

Figures~\ref{ice_profile}\&\ref{si_profile} show that the signal-to-noise ratio of our data does not allow detection of these weak features. However, the good agreement between their observed absorption profiles and those of two of the most obscured local sources: NGC4418 and IRAS08572+3915 suggests similar dust properties and source geometry.
\clearpage
\clearpage
\begin{deluxetable}{rcc ccc ccc} 
\tablecolumns{5} 
\tablewidth{0pc} 
\tabletypesize{\scriptsize}
\tablecaption{Broadband data and best-fit \con parameters with 1\,$\sigma$ uncertainties. \label{cont_table}} 
\tablehead{ 
\colhead{Source} & \colhead{$z$} & \colhead{$S_{3.6}$ [$\mu$Jy]} & \colhead{$S_{4.5}$ [$\mu$Jy]} & \colhead{$S_{5.8}$ [$\mu$Jy]} & \colhead{$S_8$ [$\mu$Jy]} & \colhead{$S_{70}$ [mJy]} & \colhead{$\alpha$} & \colhead{$\tau_{9.7}$} }
\startdata 
MIPS42 & 1.95 &   8$\pm$ 3 &  30$\pm$ 1 & 103$\pm$15 & 680$\pm$15 & 12.8$\pm$1.0 &  1.5$\pm$0.1 &  0.7$\pm$0.1 \\
MIPS78 & 2.65 &   6$\pm$ 3 &  39$\pm$ 4 &  57$\pm$24 & 268$\pm$19 &  1.5$\pm$1.3 &  1.3$\pm$0.1 &  3.4$\pm$0.5 \\
MIPS110 & 1.00 &  44$\pm$ 3 &  43$\pm$ 1 &  72$\pm$18 & 200$\pm$15 &  9.1$\pm$3.0 &  2.1$\pm$0.2 &  0.5$\pm$0.5 \\
MIPS133 & 1.00 &  57$\pm$ 6 &  54$\pm$ 4 &  55$\pm$21 & 155$\pm$17 & 13.1$\pm$1.5 &  2.3$\pm$0.1 &  0.9$\pm$0.8 \\
MIPS180 & 2.47 &  23$\pm$ 3 &  16$\pm$ 4 &  14$\pm$ 8 &  13$\pm$17 &  5.8$\pm$1.5 &  1.8$\pm$0.4 &  2.2$\pm$0.5 \\
MIPS227 & 1.87 &  72$\pm$ 3 &  82$\pm$ 3 & 111$\pm$10 & 113$\pm$19 & 10.0$\pm$2.4 &  2.9$\pm$0.2 &  0.2$\pm$0.1 \\
MIPS279 & 1.23 &  51$\pm$ 8 &  52$\pm$ 7 &  84$\pm$15 & 127$\pm$16 &  9.8$\pm$1.4 &  2.0$\pm$0.2 &  0.0$\pm$0.1 \\
MIPS283 & 0.94 & 117$\pm$ 2 & 111$\pm$ 3 & 147$\pm$23 & 131$\pm$10 & 10.2$\pm$2.1 &  1.4$\pm$0.1 &  1.5$\pm$0.9 \\
MIPS289 & 1.86 &  66$\pm$ 3 &  70$\pm$ 4 & 106$\pm$15 &  93$\pm$20 &  8.1$\pm$1.9 &  1.9$\pm$0.3 &  0.7$\pm$0.9 \\
MIPS429 & 2.15 &  18$\pm$ 6 &   6$\pm$ 4 &   4$\pm$24 &   0$\pm$20 &  3.5$\pm$1.5 &  2.7$\pm$0.2 & $>$7.3 \\
MIPS464 & 1.85 &  15$\pm$ 6 &  19$\pm$ 4 &  38$\pm$16 &  57$\pm$22 &  8.9$\pm$1.5 &  2.5$\pm$0.3 &  4.9$\pm$1.6 \\
MIPS506 &2.52 &  18$\pm$ 3 &  16$\pm$ 1 &  26$\pm$18 &  13$\pm$19 &  7.5$\pm$1.2 &  3.6$\pm$0.2 & $>$6.7 \\
MIPS8034 & 0.95 &  56$\pm$ 5 &  64$\pm$ 7 & 130$\pm$23 & 216$\pm$13 &  6.5$\pm$1.7 &  2.3$\pm$0.4 &  0.3$\pm$0.3 \\
MIPS8184 & 0.99 & 104$\pm$ 2 &  82$\pm$ 4 &  82$\pm$21 & 140$\pm$21 & 19.7$\pm$1.4 &  2.0$\pm$0.2 &  0.6$\pm$0.4 \\
MIPS8196 & 2.60 &  81$\pm$ 6 &  61$\pm$ 6 &  74$\pm$19 & 149$\pm$27 &  5.7$\pm$1.2 &  1.2$\pm$0.1 &  1.3$\pm$0.4 \\
MIPS8207 & 0.84 & 117$\pm$ 5 &  81$\pm$ 9 &  94$\pm$29 &  91$\pm$18 & 17.9$\pm$1.6 &  1.9$\pm$0.2 &  0.9$\pm$0.5 \\
MIPS8242 & 2.45 &  50$\pm$ 3 &  43$\pm$ 3 &  48$\pm$ 8 &  88$\pm$17 &  5.0$\pm$1.5 &  2.8$\pm$0.4 &  0.9$\pm$0.5 \\
MIPS8245 & 2.70 &   0$\pm$ 3 &   9$\pm$ 4 &   5$\pm$15 &   0$\pm$14 &  1.3$\pm$1.6 &  1.9$\pm$0.1 &  2.8$\pm$0.7 \\
MIPS8268 & 0.80 &  95$\pm$ 2 &  87$\pm$ 3 &  94$\pm$ 7 & 218$\pm$14 &  3.9$\pm$1.8 &  1.6$\pm$0.1 &  0.8$\pm$0.4 \\
MIPS8327 & 2.48 &  12$\pm$ 3 &   7$\pm$ 4 &   0$\pm$21 &  46$\pm$25 &  5.5$\pm$1.9 &  1.9$\pm$0.1 &  2.4$\pm$0.7 \\
MIPS8342 & 1.57 &  39$\pm$ 6 &  57$\pm$ 4 &  53$\pm$12 & 119$\pm$19 & 10.3$\pm$1.8 &  2.1$\pm$0.2 &  0.2$\pm$0.4 \\
MIPS8493 &1.80 &  39$\pm$ 6 &  48$\pm$ 4 &  62$\pm$15 &  24$\pm$14 &  4.8$\pm$1.4 &  1.4$\pm$0.7 & $>$4.7 \\
MIPS15840 & 2.30 &  18$\pm$ 2 &  25$\pm$ 1 &  62$\pm$18 & 197$\pm$13 &  2.0$\pm$1.6 &  1.6$\pm$0.2 &  0.1$\pm$0.2 \\
MIPS15880 & 1.64 &  63$\pm$ 5 &  64$\pm$ 4 &  58$\pm$16 & 118$\pm$22 & 19.5$\pm$2.1 &  2.6$\pm$0.1 &  3.9$\pm$0.4 \\
MIPS15928 & 1.52 &  39$\pm$ 3 &  51$\pm$ 4 &  58$\pm$21 & 125$\pm$26 & 12.9$\pm$2.5 &  2.6$\pm$0.2 &  0.0$\pm$0.1 \\
MIPS15949 & 2.15 &  27$\pm$ 3 &  30$\pm$ 3 &  33$\pm$ 9 &  89$\pm$11 &  7.6$\pm$1.6 &  2.3$\pm$0.2 &  0.0$\pm$0.1 \\
MIPS15958 & 1.97 &  21$\pm$ 3 &  43$\pm$ 3 &  81$\pm$24 & 181$\pm$17 &  8.0$\pm$2.8 &  1.6$\pm$0.3 &  0.5$\pm$0.4 \\
MIPS15977 & 1.85 &  57$\pm$ 3 &  61$\pm$ 4 &  69$\pm$16 & 130$\pm$22 & 15.0$\pm$1.0 &  2.2$\pm$0.2 &  0.1$\pm$0.1 \\
MIPS16030 & 0.98 &  84$\pm$ 3 &  72$\pm$ 1 &  94$\pm$16 & 132$\pm$20 &  2.3$\pm$1.6 &  1.8$\pm$0.2 &  0.4$\pm$0.4 \\
MIPS16059 & 2.43 &  17$\pm$ 3 &  19$\pm$ 4 &  33$\pm$14 &  66$\pm$15 &  6.2$\pm$2.4 &  2.1$\pm$0.1 &  2.7$\pm$0.8 \\
MIPS16080 & 2.04 &  27$\pm$ 3 &  31$\pm$ 7 &  38$\pm$24 &  71$\pm$24 &  5.5$\pm$1.7 &  1.7$\pm$0.1 &  2.1$\pm$0.5 \\
MIPS16095 & 1.81 &  21$\pm$ 2 &  21$\pm$ 4 &  24$\pm$24 &  84$\pm$14 & 11.1$\pm$2.2 &  2.2$\pm$0.2 &  0.3$\pm$0.1 \\
MIPS16113 & 1.90 &  23$\pm$ 3 &  16$\pm$ 3 &  55$\pm$15 &  35$\pm$12 &  0.9$\pm$1.5 &  1.4$\pm$0.4 &  3.3$\pm$1.3 \\
MIPS16122 & 1.97 &  17$\pm$ 3 &  34$\pm$ 9 &  31$\pm$15 & 101$\pm$15 &  3.1$\pm$1.2 &  1.9$\pm$0.4 &  1.8$\pm$0.6 \\
MIPS16144 & 2.13 &  68$\pm$ 3 &  84$\pm$ 3 & 108$\pm$24 & 105$\pm$19 &  2.2$\pm$1.2 &  2.1$\pm$0.4 &  2.0$\pm$1.4 \\
MIPS22204 & 2.08 &  20$\pm$ 9 &  39$\pm$ 6 & 132$\pm$15 & 478$\pm$ 8 & 13.2$\pm$1.1 &  1.4$\pm$0.1 &  1.6$\pm$0.2 \\
MIPS22277 & 1.77 &  69$\pm$ 3 &  67$\pm$ 4 & 142$\pm$15 & 254$\pm$ 9 & 17.5$\pm$1.3 &  1.9$\pm$0.2 &  1.4$\pm$0.3 \\
MIPS22303 & 2.34 &   9$\pm$ 3 &  10$\pm$ 4 &   0$\pm$18 & 103$\pm$13 &  7.2$\pm$2.2 &  1.7$\pm$0.1 &  2.9$\pm$0.6 \\
MIPS22404 & 0.61 & 126$\pm$ 5 &  93$\pm$ 3 & 128$\pm$16 & 117$\pm$22 & 44.3$\pm$3.0 &  2.2$\pm$0.1 &  1.5$\pm$1.4 \\
MIPS22467 & 0.80 &  59$\pm$ 9 &  40$\pm$ 4 &  31$\pm$15 & 142$\pm$25 &  9.0$\pm$1.8 &  1.8$\pm$0.1 &  0.9$\pm$0.5 \\
MIPS22482 & 1.84 &  39$\pm$ 2 &  48$\pm$ 3 &  55$\pm$18 &  93$\pm$26 & 12.1$\pm$1.1 &  1.9$\pm$0.2 &  1.2$\pm$0.4 \\
MIPS22530 &1.96 &  42$\pm$ 5 &  52$\pm$ 4 &  43$\pm$10 &  81$\pm$12 &  3.4$\pm$1.9 &  2.7$\pm$0.8 & $>$5.2 \\
MIPS22554 & 0.82 &  99$\pm$ 0 &  79$\pm$ 3 &  67$\pm$13 &  62$\pm$15 & 11.3$\pm$1.5 &  1.6$\pm$0.2 & $>$4.2 \\
MIPS22558 & 3.20 &  15$\pm$ 3 &  13$\pm$ 7 &   0$\pm$16 &  60$\pm$16 &  4.7$\pm$1.1 &  2.2$\pm$0.2 &  $>$5.6 \\
MIPS22600 & 0.86 &  93$\pm$ 2 &  58$\pm$ 7 &  46$\pm$13 &  88$\pm$20 & 10.1$\pm$1.4 &  1.7$\pm$0.2 &  2.2$\pm$3.3 \\
MIPS22651 & 1.73 &  92$\pm$ 3 & 109$\pm$ 4 &  94$\pm$15 &  99$\pm$10 &  7.0$\pm$1.0 &  2.0$\pm$0.3 &  0.7$\pm$0.5 \\
MIPS22661 & 1.75 &  36$\pm$ 2 &  30$\pm$ 1 &  12$\pm$ 8 &  76$\pm$19 &  2.3$\pm$2.4 &  2.3$\pm$0.2 &  0.0$\pm$0.1 \\
MIPS22699 & 2.59 &   9$\pm$ 5 &  18$\pm$ 4 &  17$\pm$19 & 105$\pm$13 &  2.5$\pm$2.3 &  1.2$\pm$0.2 &  1.2$\pm$0.7 \\
\enddata
\end{deluxetable}

\begin{deluxetable}{rccccc}
\tablecolumns{8}
\tablewidth{0pc}
\tabletypesize{\scriptsize}
\tablecaption{\label{lums_table} Luminosities with 1\,$\sigma$ uncertainties. Upper limits are 3\,$\sigma$. }
\tablehead{\colhead{Source}  & \colhead{$\log(L_{1.6}/L_{\odot})$} & \colhead{$\log(L_{5.8}/L_{\odot})$} & \colhead{$\log(L_{14}/L_{\odot})$} & \colhead{$\log(L_{30}/L_{\odot})$} & \colhead{$\log(L_{PAH,7.7}/L_{\odot})$} }  
 \startdata
MIPS42 & --- & 12.51$\pm$0.02 & 12.68$\pm$0.03 & --- & 10.60$\pm$0.20 \\
MIPS78 & --- & 12.66$\pm$0.02 & 12.76$\pm$0.04 & --- & $<$10.56 \\
MIPS110 & 10.81$\pm$0.10 & 11.21$\pm$0.07 & 11.64$\pm$0.02 & 11.84$\pm$0.52 &  9.73$\pm$0.19 \\
MIPS133 & 10.81$\pm$0.14 & 10.97$\pm$0.05 & 11.45$\pm$0.03 & 11.71$\pm$0.27 & $<$ 9.70 \\
MIPS180 & 11.10$\pm$0.70 & 12.32$\pm$0.03 & 12.57$\pm$0.07 & --- & 10.92$\pm$0.10 \\
MIPS227 & 11.53$\pm$0.10 & 11.68$\pm$0.06 & 12.39$\pm$0.02 & --- & 10.26$\pm$0.18 \\
MIPS279 & 10.89$\pm$0.35 & 10.19$\pm$0.08 & 11.57$\pm$0.02 & 11.92$\pm$0.34 &  9.65$\pm$0.20 \\
MIPS283 & 11.12$\pm$0.03 & 10.85$\pm$0.08 & 11.01$\pm$0.04 & 11.48$\pm$0.72 & 10.38$\pm$0.03 \\
MIPS289 & 11.47$\pm$0.15 & 11.47$\pm$0.23 & 11.80$\pm$0.22 & --- & 11.14$\pm$0.05 \\
MIPS429 & --- & 11.66$\pm$0.09 & 12.22$\pm$0.03 & --- & 10.63$\pm$0.09 \\
MIPS464 & --- & 11.53$\pm$0.06 & 12.16$\pm$0.03 & --- & 10.09$\pm$0.19 \\
MIPS506 & 11.17$\pm$0.35 & 11.82$\pm$0.20 & 12.74$\pm$0.01 & --- & 11.11$\pm$0.13 \\
MIPS8034 & 10.75$\pm$0.20 & 11.21$\pm$0.08 & 11.66$\pm$0.02 & 11.58$\pm$0.48 & $<$ 9.59 \\
MIPS8184 & 10.88$\pm$0.08 & 10.94$\pm$0.06 & 11.32$\pm$0.02 & 11.87$\pm$0.23 & 10.44$\pm$0.03 \\
MIPS8196 & 11.74$\pm$0.24 & 12.46$\pm$0.02 & 12.53$\pm$0.02 & --- & 11.09$\pm$0.08 \\
MIPS8207 & 10.89$\pm$0.12 & 10.71$\pm$0.07 & 11.06$\pm$0.04 & 11.59$\pm$0.32 & 10.20$\pm$0.03 \\
MIPS8242 & 11.51$\pm$0.21 & 12.07$\pm$0.04 & 12.74$\pm$0.04 & --- & 10.70$\pm$0.19 \\
MIPS8245 & --- & 12.23$\pm$0.08 & 12.55$\pm$0.03 & --- & 10.41$\pm$0.18 \\
MIPS8268 & 10.79$\pm$0.05 & 10.82$\pm$0.05 & 11.04$\pm$0.02 & 11.19$\pm$0.80 & $<$ 9.05 \\
MIPS8327 & --- & 12.07$\pm$0.07 & 12.40$\pm$0.04 & --- & 10.62$\pm$0.21 \\
MIPS8342 & 11.00$\pm$0.35 & 11.50$\pm$0.08 & 11.93$\pm$0.02 & 12.28$\pm$0.36 & 10.52$\pm$0.10 \\
MIPS8493 & 11.27$\pm$0.27 & 11.38$\pm$0.23 & 11.49$\pm$0.22 & --- & 10.93$\pm$0.06 \\
MIPS15840 & 11.22$\pm$0.20 & 12.28$\pm$0.03 & 12.50$\pm$0.04 & --- & $<$10.29 \\
MIPS15880 & 11.28$\pm$0.18 & 11.94$\pm$0.03 & 12.51$\pm$0.01 & --- & 10.60$\pm$0.09 \\
MIPS15928 & 10.95$\pm$0.23 & 11.44$\pm$0.05 & 12.04$\pm$0.02 & 12.36$\pm$0.37 & 10.69$\pm$0.05 \\
MIPS15949 & 11.16$\pm$0.35 & 11.94$\pm$0.04 & 12.45$\pm$0.03 & --- & 10.71$\pm$0.08 \\
MIPS15958 & --- & 11.95$\pm$0.05 & 12.19$\pm$0.04 & --- & 10.29$\pm$0.20 \\
MIPS15977 & 11.36$\pm$0.15 & 11.82$\pm$0.05 & 12.28$\pm$0.02 & --- & $<$10.24 \\
MIPS16030 & 10.98$\pm$0.07 & 10.93$\pm$0.07 & 11.22$\pm$0.06 & 11.39$\pm$0.75 &  9.91$\pm$0.08 \\
MIPS16059 & 11.17$\pm$0.57 & 12.05$\pm$0.04 & 12.46$\pm$0.02 & --- & 10.76$\pm$0.09 \\
MIPS16080 & 11.19$\pm$0.42 & 11.93$\pm$0.04 & 12.18$\pm$0.02 & --- & 10.20$\pm$0.17 \\
MIPS16095 & 10.85$\pm$0.32 & 11.69$\pm$0.05 & 12.12$\pm$0.02 & --- & $<$10.16 \\
MIPS16113 & --- & 11.77$\pm$0.08 & 11.88$\pm$0.16 & --- & 10.80$\pm$0.08 \\
MIPS16122 & --- & 11.81$\pm$0.06 & 12.13$\pm$0.09 & --- & 10.41$\pm$0.21 \\
MIPS16144 & 11.67$\pm$0.10 & 11.59$\pm$0.28 & 12.01$\pm$0.14 & --- & 11.20$\pm$0.06 \\
MIPS22204 & --- & 12.50$\pm$0.02 & 12.63$\pm$0.01 & --- & 11.01$\pm$0.10 \\
MIPS22277 & 11.36$\pm$0.13 & 12.03$\pm$0.08 & 12.36$\pm$0.05 & --- & 10.44$\pm$0.19 \\
MIPS22303 & --- & 12.32$\pm$0.06 & 12.57$\pm$0.03 & --- & $<$10.50 \\
MIPS22404 & 10.84$\pm$0.04 & 10.32$\pm$0.07 & 10.77$\pm$0.04 & 11.52$\pm$0.33 & --- \\
MIPS22467 & 10.52$\pm$0.26 & 10.57$\pm$0.05 & 10.86$\pm$0.02 & 11.31$\pm$0.62 & --- \\
MIPS22482 & 11.19$\pm$0.15 & 11.71$\pm$0.08 & 12.04$\pm$0.14 & --- & 10.86$\pm$0.13 \\
MIPS22530 & 11.33$\pm$0.26 & 11.48$\pm$0.25 & 12.10$\pm$0.15 & --- & 11.02$\pm$0.10 \\
MIPS22554 & 10.91$\pm$0.03 & 10.50$\pm$0.09 & 10.71$\pm$0.08 & 11.29$\pm$0.57 &  9.96$\pm$0.04 \\
MIPS22558 & --- & 12.49$\pm$0.02 & 12.88$\pm$0.01 & --- & 10.96$\pm$0.12 \\
MIPS22600 & 10.83$\pm$0.10 & 10.59$\pm$0.10 & 10.83$\pm$0.08 & 11.34$\pm$0.53 & 10.04$\pm$0.05 \\
MIPS22651 & 11.50$\pm$0.10 & 11.50$\pm$0.15 & 11.86$\pm$0.05 & --- & 10.77$\pm$0.07 \\
MIPS22661 & 10.94$\pm$0.21 & 11.44$\pm$0.07 & 11.93$\pm$0.04 & --- & 10.21$\pm$0.21 \\
MIPS22699 & --- & 12.12$\pm$0.04 & 12.17$\pm$0.08 & --- & $<$10.34 \\
\enddata
\end{deluxetable}

\begin{deluxetable}{cccccccc}
\tablecolumns{8}
\tablewidth{0pc}
\tabletypesize{\scriptsize}
\tablecaption{\label{pah_table} PAH strength with 1\,$\sigma$ uncertainties. Upper limits are 2\,$\sigma$. }
\tablehead{\colhead{Source} & \colhead{EW6.2} & \colhead{EW7.7} & \colhead{EW11.3} & \colhead{$F_{6.2}$} & \colhead{$F_{7.7}$} & \colhead{$F_{11.3}$} &  \\
\colhead{} &  \colhead{\um} & \colhead{\um} & \colhead{\um} & \colhead{$10^{-15}$ergs~s$^{-1}$\,cm$^{-2}$} & \colhead{$10^{-15}$ergs~s$^{-1}$\,cm$^{-2}$} & \colhead{$10^{-15}$ergs~s$^{-1}$\,cm$^{-2}$} & \colhead{} }
 \startdata
MIPS42 & $<$0.08 &  0.08$\pm$ 0.04 & $<$0.04 & $<$6.8 &  6.2$\pm$2.9 & $<$2.2 & \\
MIPS78 & $<$0.04 & $<$0.06 & --- & $<$2.6 & $<$2.8 & --- \\
MIPS110 & $<$0.26 &  0.18$\pm$ 0.08 & $<$0.08 & $<$6.0 &  4.4$\pm$1.9 & $<$2.0 & \\
MIPS133 & $<$0.30 & $<$0.30 & $<$0.26 & $<$5.4 & $<$5.2 & $<$3.4 & \\
MIPS180 &  0.08$\pm$0.04 &  0.26$\pm$0.07 & --- & $<$2.6 &  7.3$\pm$1.6 & --- & \\
MIPS227 & $<$0.22 &  0.17$\pm$0.08 &  0.08$\pm$0.04 & $<$3.4 &  3.2$\pm$1.3 & $<$2.0 & \\
MIPS279 & --- &  0.17$\pm$0.08 &  0.25$\pm$0.08 & --- &  2.2$\pm$1.0 &  3.3$\pm$1.1 & \\
MIPS283 & --- &  2.66$\pm$0.33 &  1.85$\pm$0.37 & --- & 22.8$\pm$1.6 &  9.9$\pm$1.6 & \\
MIPS289 &  0.48$\pm$0.14 &  2.81$\pm$0.23 &  0.82$\pm$0.13 &  4.4$\pm$1.4 & 24.3$\pm$2.7 &  5.4$\pm$1.7 \\
MIPS429 & $<$0.30 &  0.41$\pm$ 0.09 & $<$2.4 & $<$3.2 &  5.2$\pm$1.0 & $<$2.6 & \\
MIPS464 & $<$0.12 &  0.51$\pm$ 0.10 & $<$0.32 & $<$1.4 &  6.9$\pm$1.2 & $<$1.2 & \\
MIPS506 &  0.30$\pm$0.14 &  0.87$\pm$0.36 & --- &  3.4$\pm$1.7 & 10.8$\pm$3.3 & --- & \\
MIPS8034 & $<$0.26 & $<$0.22 &  0.23$\pm$0.10 & $<$5.8 & $<$5.4 &  5.8$\pm$2.5 \\
MIPS8184 & --- &  1.83$\pm$0.11 &  1.08$\pm$0.13 & --- & 22.8$\pm$1.3 & 11.7$\pm$1.3 & \\
MIPS8196 & $<$0.06 &  0.31$\pm$0.08 & --- & $<$2.0 &  9.5$\pm$1.7 & --- & \\
MIPS8207 & --- &  2.13$\pm$0.21 &  0.77$\pm$0.13 & --- & 20.0$\pm$1.5 &  6.4$\pm$1.1 & \\
MIPS8242 & $<$0.12 &  0.21$\pm$0.10 & --- & $<$2.4 &  4.5$\pm$2.0 & --- & \\
MIPS8245 & $<$0.16 &  0.09$\pm$0.04 & --- & $<$3.4 &  1.8$\pm$0.8 & --- & \\
MIPS8268 & --- & $<$0.12 & $<$0.36 & --- & $<$1.6 & $<$3.6 & \\
MIPS8327 & $<$0.14 & $<$0.30 & --- & $<$2.4 & $<$3.6 & --- \\
MIPS8342 & $<$0.16 &  0.58$\pm$ 0.14 &  0.11$\pm$ 0.04  & $<$2.2 &  8.4$\pm$2.0 &  1.6$\pm$ 0.5 & \\
MIPS8493 &  0.32$\pm$0.13 &  2.72$\pm$0.43 &  2.72$\pm$0.52 &  2.4$\pm$1.1 & 16.4$\pm$2.2 &  4.2$\pm$1.0 \\
MIPS15840 & $<$0.10 & $<$0.06 & $<$0.22 & $<$3.2 & $<$2.0 & $<$5.0 & \\
MIPS15880 &  0.06$\pm$0.03 &  0.23$\pm$0.05 & $<$0.06 & $<$2.2 &  8.9$\pm$1.9 & $<$1.0 & \\
MIPS15928 &  0.33$\pm$0.07 &  0.89$\pm$0.11 &  0.20$\pm$0.03 &  4.7$\pm$1.0 & 14.2$\pm$1.7 &  4.0$\pm$0.6 \\
MIPS15949 & $<$0.14 &  0.31$\pm$ 0.06 & $<$0.12 & $<$2.8 &  6.4$\pm$1.2 & $<$2.6 & \\
MIPS15958 & $<$0.14 &  0.14$\pm$ 0.06 & $<$0.12 & $<$3.0 &  3.0$\pm$1.4 & $<$1.8 & \\
MIPS15977 &  0.18$\pm$0.07& $<$0.14 &  0.06$\pm$0.03 &  3.7$\pm$1.4& $<$3.0 &  1.3$\pm$0.7 \\
MIPS16030 & --- &  0.60$\pm$0.12 &  0.58$\pm$0.11 & --- &  7.0$\pm$1.3 &  5.6$\pm$1.1 & \\
MIPS16059 & $<$0.14 &  0.29$\pm$0.07 & --- & $<$2.6 &  5.3$\pm$1.2 & --- & \\
MIPS16080 & $<$0.14 &  0.11$\pm$ 0.05 & $<$0.20 & $<$2.8 &  2.2$\pm$ 0.9 & $<$1.8 & \\
MIPS16095 & $<$0.12 & $<$0.16 & $<$0.06 & $<$2.4 & $<$3.2 & $<$1.0 & \\
MIPS16113 & $<$0.16 &  0.79$\pm$ 0.16 & $<$0.38 & $<$2.6 & 10.6$\pm$ 2.1 & $<$1.6 & \\
MIPS16122 & $<$0.16 &  0.25$\pm$ 0.11 &  0.19$\pm$ 0.09  & $<$2.8 &  3.9$\pm$ 1.9 &  1.7$\pm$ 0.8 & \\
MIPS16144 & $<$0.32 &  2.50$\pm$ 0.26 &  0.50$\pm$ 0.24 & $<$2.8 & 20.2$\pm$ 2.8 & $<$2.6 & \\
MIPS22204 & $<$0.08 &  0.22$\pm$ 0.05 & $<$0.10 & $<$7.8 & 13.6$\pm$ 3.1 & $<$2.8 & \\
MIPS22277 & $<$0.12 &  0.15$\pm$ 0.07 & $<$0.08 & $<$4.2 &  5.5$\pm$ 2.4 & $<$2.8 & \\
MIPS22303 & $<$0.12 & $<$0.10 & --- & $<$4.6 & $<$3.2 & --- \\
MIPS22404 & --- & --- &  1.68$\pm$0.16 & --- & --- & 12.7$\pm$1.2 & \\
MIPS22467 & --- & --- & $<$0.26 & --- & --- & $<$2.2 & \\
MIPS22482 & $<$0.16 &  0.90$\pm$ 0.15 & $<$0.12 & $<$2.6 & 13.0$\pm$3.8 & $<$3.2 & \\
MIPS22530 &  0.39$\pm$0.15 &  1.85$\pm$0.32 &  0.91$\pm$0.29 &  3.4$\pm$1.3 & 16.2$\pm$3.8 &  3.4$\pm$1.5 \\
MIPS22554 & --- &  2.09$\pm$0.31 &  4.59$\pm$1.13 & --- & 12.0$\pm$1.1 &  9.1$\pm$1.6 & \\
MIPS22558 & $<$0.04 &  0.16$\pm$0.08 & --- & $<$1.2 &  4.3$\pm$1.2 & --- & \\
MIPS22600 & --- &  1.94$\pm$0.32 &  2.38$\pm$0.63 & --- & 13.0$\pm$1.4 &  8.6$\pm$1.7 & \\
MIPS22651 &  0.28$\pm$0.10 &  1.12$\pm$0.18 &  0.21$\pm$0.06 &  3.2$\pm$1.2 & 12.4$\pm$2.0 &  2.0$\pm$0.5 \\
MIPS22661 & $<$0.28 &  0.32$\pm$ 0.16 &  0.17$\pm$ 0.06  & $<$2.6 &  3.1$\pm$1.5 &  1.8$\pm$0.6 & \\
MIPS22699 & $<$0.12 & $<$0.12 & --- & $<$2.0 & $<$1.8 & --- \\
\enddata
\end{deluxetable}

\clearpage

\clearpage

\end{document}